%% file: main.tex
\documentclass[journal=jpcc,manuscript=article]{achemso}
\usepackage[utf8]{inputenc}
\usepackage{amsmath}
\usepackage{amsfonts}
\usepackage{amssymb}
\usepackage{hyperref}
\usepackage{tikz,lipsum,lmodern}
\usepackage[most]{tcolorbox}
\usepackage{siunitx}
\usepackage{xcolor}
\usepackage{multirow}
\usepackage{array}
\usepackage{tabularx}
\usepackage{booktabs}
\usepackage{dsfont}
\usepackage{caption}
\usepackage{subcaption}
\usepackage{cleveref}
\usepackage{float}

\DeclareSIUnit\angstrom{\text{Å}}

\newcommand{\bra}[1]{\langle #1 |}
\newcommand{\ket}[1]{|#1\rangle}
\newcommand{\braket}[2]{\langle #1 | #2 \rangle}
\newcommand{\comm}[2]{[\,#1, #2\,]}
\newcommand{\cre}[1]{a^\dagger_{#1}}
\newcommand{\ann}[1]{a_{#1}}

\newcommand{\vacbra}{\bra{\text{vac}}}
\newcommand{\vacket}{\ket{\text{vac}}}

\newcommand{\tcre}[1]{\tilde a^\dagger_{#1}}
\newcommand{\tann}[1]{\tilde a_{#1}}

\newcommand{\half}{\frac{1}{2}}
\newcommand{\eps}{\varepsilon}
\newcommand{\Tr}[2]{\mathrm{Tr}_{#1}\{#2\}}
\newcommand{\ddt}{\frac{d}{d t}}

\newcommand{\systa}{\text S}
\newcommand{\systb}{\text E}
\newcommand{\pproj}{\mathcal{P}}
\newcommand{\qproj}{\mathcal{Q}}

\SectionNumbersOn

\title{Deriving the Redfield Equation for Electronically Open Molecules}

\author{Bendik Støa Sannes}
\affiliation{Department of Chemistry, The Norwegian University of Science and Technology, Trondheim, Norway}
\author{Jacob Pedersen$^{\dagger}$}
\affiliation{Department of Chemistry, Technical University of Denmark, Kongens Lyngby, Denmark}
\author{Ida-Marie Høyvik}
\affiliation{Department of Chemistry, The Norwegian University of Science and Technology, Trondheim, Norway}
\email{ida-marie.hoyvik@ntnu.no}

\begin{document}

\begin{abstract}
We introduce a formalism to describe fractional charging of a molecule due to interactions with its environment. The interactions which induce fractional charging are contained in the Hamiltonian of the full system (molecule and environment). Such interactions can be singled out by expressing the Hamiltonian in a local spin orbital basis, and they are the main focus of this work. A reduced density operator for the molecule is derived starting from the Liouville-von Neumann equation for the full system by employing an explicitly constructed projection superoperator. By treating the molecule as an electronically open quantum system, we obtain a Redfield equation where the environment is included approximately. Phenomenological broadening of energy levels is included to mimic finite lifetimes of electronic states. The populations of the reduced density operator determine the mixture of different redox states and, hence, the fractional charging of the molecule. To illustrate the formalism, we use benzene physisorbed on a graphene sheet as a toy model. The work presented in this paper constitutes an initial step toward understanding molecules as electronically open quantum systems. 
\end{abstract}

\section{Introduction}
\label{sec:introduction}

State-of-the-art electronic structure models may describe a molecule interacting with its environment. Numerous approaches are developed to include such environmental effects, either quantum mechanically\cite{hofener2012wfinwf,saether2017mlhf,myhre2014mlcc,jacob2014subsystem,lee2019projection,manby2012simple} or classically.\cite{senn2009qm,cappelli2016integrated,tzeliou2022qmmmreview} A common feature of these approaches is that they only include interactions that conserve the number of electrons in the molecule and environment separately. Hence, the molecule is treated as an electronically closed system.
However, if a full quantum-mechanical description of both the molecule and the environment is carried out, electrons are not strictly enforced to belong to either subsystem. Consequently, a full quantum-mechanical description allows for electron transfer between subsystems. The electron transfer may range from integer transfer (redox reaction) to partial transfer (fractional charging). Fractional charging is the manifestation of electrons being shared to some extent between the subsystems, and while integer electron transfer is widely studied\cite{marcus1964chemical,marcus1985electron,newton1991quantum,fletcher2010theory,blumberger2015recent}, fractional charging is largely ignored. 

We will describe the fractional charging of a molecule due to the interaction with its environment. Through an analysis of the Hamiltonian of the full system in a basis of spin orbitals localized to either of the subsystems, the terms inducing fractional charging are identified. From these terms, an effective interaction operator is constructed. Our analysis of the Hamiltonian directly shows which one- and two-electron integrals are contained in the effective interaction strength. 
We note that the effective operator is of the same form as in electron transport models\cite{dzhioev2012nonequilibrium,Galperin:2012aa,elenewski2017communication,Philbin_Levy_Narang_Dou_2021,Naaman:2022aa}.
To describe the effect of the interaction on the molecule, we will use the density operator formalism and the theory of open quantum systems.\cite{breuer_2002_theory_of_open_quantum_systems,may_2011_charge_and_energy_transfer_dynamics_in_molecular_systems,nitzan_2013_chemical_dynamics_in_condensed_phases,rivas_2011_open_quantum_systems_an_introduction,stefanucci_2013_nonequilibrium_many_body_theory} The theory of open quantum systems aims to describe the time evolution of a system in the presence of an environment, where only the environment's effect on the target system is included approximately. Previously, the theory has been used to model numerous phenomena, such as vibrational relaxation for spectroscopy\cite{Yan:1988aa,Khalil:2004aa,Galperin:2008aa} and energy transfer in electron transfer processes.\cite{jean1992application,Leegwater:1997aa,Jeske:2015aa,Storm:2019aa,Pedersen_2022_redfield_propagation_photoinduced_ET}

The advantage of using density operators is that they can describe irreversible processes not attainable when using the electronic wave function. Examples are dissipation and decoherence\cite{weiss_2012_quantum_dissipative_systems,blum2012density,TheConceptofDecoherence_2013}, and the extent of these effects depends on the interaction between the molecule and the environment.\cite{breuer_2002_theory_of_open_quantum_systems} Furthermore, the density operator allows for direct access to the state populations.
The time development of the reduced density operator describing only the molecule explicitly is derived from the Liouville-von Neumann equation for the full system. This is achieved by using the projection approach of Nakajima and Zwanzig\cite{nakajima1958quantum,Zwanzig_1960_projection_operators}, resulting in two coupled equations. In the absence of the interaction operator, the two equations decouple and correspond to the Liouville-von Neumann equations for the separate subsystems. Hence, the interaction couples the electronic degrees of freedom in the molecule to the electronic degrees of freedom in the environment. 

The coupled equations resulting from the Nakajima-Zwanzig approach are as challenging to solve as the Liouville-von Neumann equation for the full system. To simplify, the Born and Markov approximations are typically introduced.\cite{breuer_2002_theory_of_open_quantum_systems,nitzan_2013_chemical_dynamics_in_condensed_phases,may_2011_charge_and_energy_transfer_dynamics_in_molecular_systems} These approximations assume that the environment is large and not significantly altered by the interaction, and that the environment's relaxation time is short. Additionally, assuming that the molecule-environment interaction is weak leads to the Redfield equation. Applying the secular approximation on top of this yields a master equation of Lindblad form, ensuring complete positivity of the dynamics\cite{lindblad1976generators,gorini1976completely,manzano2020short,hartmann2020accuracy,tupkari_2022_limitations_of_lindblad_equations}. The Redfield and Lindblad equations have been extensively used in quantum optics\cite{mukamel1995principles,campaioli2024quantum}, and for electron transport and charge separation in molecular system.\cite{jean1992application,novoderezhkin2004coherent,kilin2000electron,novoderezhkin2007mixing,dzhioev2011super,may_2011_charge_and_energy_transfer_dynamics_in_molecular_systems,dzhioev2012nonequilibrium,elenewski2017communication,Pedersen_2022_redfield_propagation_photoinduced_ET}

We derive the Redfield equation using the exact eigenstates of the molecular electronic Hamiltonian for the non-interacting molecule and environment. Within a given one-electron basis, the exact eigenstates are the full configuration interaction (FCI) states.\cite{Shavitt1977,helgaker2014_mest} It is important to note that due to the form of the interaction operator, we need to consider states with different number of electrons. Since FCI calculations are only applicable to very small molecular systems, we use Slater determinants as the many-electron basis for the reduced density operator in the derived form of the Redfield equation. The determinants do not provide a quantitative description of the molecule, but serve the purpose of illustrating the presented formalism. From propagating the equation to a steady state, the populations of the determinants are obtained. 
The average number of electrons in the molecule may be fractional due to mixing of determinants with different number of electrons. In this formalism, fractional charging can be calculated as the deviation from the number of electrons of the isolated molecule. 

In standard electronic structure theory, fractional orbital occupations have been extensively used to recover static correlation for isolated molecules.\cite{bogolyubov1959compensation,slater_1969_hhf,abdulnur_1972_gchf,zerner_1989_cahf,perdew_1982_fons_dft,ullrich_2001_ensembleDFT,lee_2022_recent_advanced_in_edft,mussard_2017_fon_dft}. On the other hand, the particle-breaking Hartree-Fock (PBHF) model\cite{Matveeva_2023_PBHF,matveeva2024particle} is developed to describe the molecule as an electronically open quantum system. The PBHF wave function is obtained by using the time-independent Schrödinger equation and the variational principle. The interaction operator is parametrized prior to the optimization, and PBHF therefore represents a convenient approach to capture fractional charging effects. However, irreversible processes due to the interactions cannot be represented by the wave function.

The paper is organized as follows. In Section \ref{sec:theory}, we present the subsystem analysis of the Hamiltonian in the local spin-orbital basis and use FCI states of the non-interacting subsystems to construct the density operator for the full system. Next, we show the derivation of the Redfield equation for the populations of the reduced density operator. This involves the use of an explicitly constructed projection superoperator. In Section \ref{sec:results_and_discussion}, we present an illustrative example by using a benzene molecule physisorbed on graphene as a toy model. We further discuss the developments necessary to obtain an appropriate electronic description of the molecule. Finally, Section \ref{sec:conclusions} provides a summary and concluding remarks.

\section{Theory}
\label{sec:theory}

We start from the full configuration interaction (FCI) solution to the electronic problem of 
a closed system, as this provides eigenstates of the Hamiltonian within the given one-electron basis. Next, we partition the closed system into two interacting subsystems, the molecule and its environment. The goal is to obtain a reduced description of the molecule while retaining the components that induce fractional charging. We construct the density operator for the full system (molecule and environment) starting from the exact solutions of the isolated subsystems. We then present the steps involved in the derivation of the Redfield equation, which gives the time evolution of the reduced density operator. Finally, we briefly discuss how to represent properties of the environment and the effect of energy level broadening.

\subsection{Electronic structure theory for closed systems}
\label{sec:mest_for_closed_systems}
The time-independent electronic Schrödinger equation is given by
\begin{align}
    \hat H\Psi = E\Psi
\end{align}
where $\hat H$ is the molecular electronic Hamiltonian, $\Psi$ is the electronic wave function and $E$ is the energy. 
The Hamiltonian in second quantization, excluding nuclear repulsion, is given by
\begin{align}
\label{eq:electronic_hamiltonian}
    \hat H = \sum_{PQ}h_{PQ}\cre{P}\ann{Q} + \half\sum_{PQRS}g_{PQRS}\cre{P}\cre{R}\ann{S}\ann{Q}
\end{align}
where $\cre{P}$ and $\ann{P}$ are the creation and annihilation operators for spin-orbital $\phi_P$, respectively. The set of creation and annihilation operators for the spin orbitals obey the anticommutation rules $\comm{\cre{P}}{\cre{Q}}_+=\comm{\ann{P}}{\ann{Q}}_+=0$, and $\comm{\cre{P}}{\ann{Q}}_+=\delta_{PQ}$. Further, we have
\begin{align}
    h_{PQ} &= \int \phi_P^*(\mathbf{x}_1)\Bigl(-\half\nabla^2-\sum_I\frac{Z_I}{|\mathbf{r}_1-\mathbf{R}_I|}\Bigr)\phi_Q(\mathbf{x}_1)d\mathbf{x}_1\\
    g_{PQRS} &= \int \int \phi_P^*(\mathbf{x}_1)\phi_Q(\mathbf{x}_1)\frac{1}{|\mathbf{r}_1-\mathbf{r}_2|}\phi_R^*(\mathbf{x}_2)\phi_S(\mathbf{x}_2)d\mathbf{x}_1d\mathbf{x}_2\label{eq:two_el_integral}
\end{align}
The integration variable $\mathbf{x}_i$ includes spatial and spin coordinates of electron $i$, 
$Z_I$ is the charge of nucleus $I$, and $\mathbf{r}_i$ and $\mathbf{R}_I$ are electron and nuclear coordinates, respectively. The Hamiltonian in eq \ref{eq:electronic_hamiltonian} represents a closed molecular system, and it commutes with the electron number operator $\hat N=\sum_P\cre{P}\ann{P}$. In other words, the number of electrons is conserved for the closed system.

The exact solution within a chosen atomic orbital basis is given by the FCI wave function,\cite{Shavitt1977,helgaker2014_mest} 
\begin{align}
\label{eq:fci_state}
    \ket{\Psi_k} = \sum_{I}C_{Ik}\ket I
\end{align}
The sum over $I$ includes all determinants for distributing $N$ electrons among $M$ spin orbitals, with corresponding expansion coefficients $C_{Ik}$. The FCI states are orthogonal and may be normalized,
\begin{align}
\begin{split}
    \braket{\Psi_k}{\Psi_l} = \delta_{kl}
\end{split}
\end{align}
The variational minimization of the energy with respect to the expansion coefficients amounts to diagonalizing $\mathbf{H}$, i.e., solving the FCI eigenproblem
\begin{align}
    \mathbf{HC}_k=E_k\mathbf{C}_k
\end{align}
where $E_k$ is the energy of the $k$-th state, $\ket{\Psi_k}$, and the elements $H_{IJ}=\bra{I} \hat H \ket{J}$ are evaluated using the Slater-Condon rules.

\subsection{Dividing the closed system into two interacting subsystems}
\label{sec:dividing_system}

We now consider an overall closed system, which we partition into a region of interest (the molecular system, $\systa$) and everything else (the environment, $\systb$). Henceforth, we will use the term environment for everything that is not treated explicitly. We further assume that we can transform the set of spin orbitals of the full system into a set where each spin orbital can be assigned to either the system or the environment. As long as there is no covalent bond between the subsystems, we know such localized spin orbitals exist.\cite{hoyvik2016characterization} 
We note that this does not mean that the spin orbitals are all spatially localized, as they may be delocalized across either the molecule or the environment.

From now on, we will denote spin orbitals belonging to the molecule's orbital space with unbarred indices $\{\phi_p\}$, and orbitals belonging to the environment with barred indices $\{\phi_{\bar p}\}$, such that the set of all spin orbitals $\{\phi_P\} = \{\phi_p\}\cup\{\phi_{\bar p}\}$. We also use this distinction of indices for the second quantized operators. We may now separate the Hamiltonian in eq \ref{eq:electronic_hamiltonian} into three parts,
\begin{align}
    \hat H = \hat H_\systa + \hat H_\systb + \hat H_{\systa\systb}
\end{align}
with
\begin{align}
    \hat H_\systa &= \sum_{pq}h_{pq}\cre{p}\ann{q} + \half\sum_{pqrs}g_{pqrs}\cre{p}\cre{r}\ann{s}\ann{q}\\
    \hat H_\systb &= \sum_{\bar p\bar q}h_{\bar p\bar q}\cre{\bar p}\ann{\bar q} + \half\sum_{\bar p\bar q\bar r\bar s}g_{\bar p\bar q\bar r\bar s}\cre{\bar p}\cre{\bar r}\ann{\bar s}\ann{\bar q}\\
    \hat H_{\systa\systb} &= \hat H - \hat H_\systa - \hat H_\systb
\end{align}
$\hat H_\systa$ and $\hat H_\systb$ describe the two subsystems separately, and they commute. $\hat H_{\systa\systb}$ contains all interactions between the molecule and the environment. We may also divide the electron number operator into a number operator for the molecule and for the environment, $\hat N= \hat N_\systa + \hat N_\systb$. The subsystem number operators are
\begin{align}
    \hat N_\systa &=\sum_p\cre{p}\ann{p}\\
    \hat N_\systb &= \sum_{\bar p}\cre{\bar p}\ann{\bar p}
\end{align}
The molecule and environment number operators commute with their corresponding Hamiltonians, i.e., $\comm{\hat H_\systa}{\hat N_\systa} = \comm{\hat H_\systb}{\hat N_\systb} = 0$. In contrast, although the interaction Hamiltonian $\hat H_{\systa\systb}$ commutes with $\hat N$, only parts of it commute with $\hat N_\systa$ and $\hat N_\systb$ separately. We will therefore separate $\hat H_{\systa\systb}$ into two components, 
\begin{align}
     \hat H_{\systa\systb}=\hat H_{\systa\systb}^\text{pc}+\hat H_{\systa\systb}^\text{pb}
\end{align}
where $\hat H_{\systa\systb}^\text{pc}$ and $\hat H_{\systa\systb}^\text{pb}$ contain the particle-conserving and particle-breaking terms, respectively. By particle-breaking, we mean non-particle-conserving.\cite{Matveeva_2023_PBHF,matveeva2024particle}
These properties follow from the commutation relations
\begin{align}
    \comm{\hat H_{\systa\systb}^\text{pc}}{\hat N_\systa} &= 0,\qquad
    \comm{\hat H_{\systa\systb}^\text{pc}}{\hat N_\systb} = 0\\
    \comm{\hat H_{\systa\systb}^\text{pb}}{\hat N_\systa} &\neq 0, \qquad
    \comm{\hat H_{\systa\systb}^\text{pb}}{\hat N_\systb} \neq  0
\end{align}
Interactions contained in $\hat H^\text{pc}_{\systa\systb}$ may change the energies of the subsystems, but the respective particle numbers in each subsystem are unchanged. Numerous approaches\cite{hofener2012wfinwf,saether2017mlhf,myhre2014mlcc,jacob2014subsystem,lee2019projection,manby2012simple,senn2009qm,cappelli2016integrated,tzeliou2022qmmmreview} include these types of interactions and will therefore not be the focus of this work. However, interactions contained in $\hat H^\text{pb}_{\systa\systb}$ results in electron number fluctuations which may induce fractional charging. Here, we define fractional charging by
\begin{align}
\label{eq:fractional_charging}
    \delta N_\systa = N_{\systa,0} - \langle \hat N_\systa \rangle
\end{align}
where $N_{\systa,0}$ is the number of electrons in the isolated system, and $\langle\hat N_\systa\rangle$ is the average number of electrons in the molecule upon interaction with the environment.

\subsubsection{The particle-breaking interaction operator}

We will now reduce $\hat H_{\systa\systb}^\text{pb}$ to an approximate effective interaction operator $\hat V$. All terms of $\hat H_{\systa\systb}^\text{pb}$ are
\begin{align}
\begin{split}
    \hat H^\text{pb}_{\systa\systb}&= \sum_{p\bar q}\bigl(h_{p\bar q}\cre{p}\ann{\bar q} + h_{p\bar q}^*\cre{\bar q}\ann{p}\bigr)
    + \sum_{p\bar qrs}\bigl(g_{p\bar qrs}\cre{p}\cre{r}\ann{s}\ann{\bar q}
    +  g_{p\bar qrs}^*\cre{\bar q}\cre{s}\ann{r}\ann{p} \bigr) \\
    &+ \half\sum_{p\bar qr\bar s}\bigl(g_{p\bar qr\bar s}\cre{p}\cre{r}\ann{\bar s}\ann{\bar q} + g_{p\bar qr\bar s}^*\cre{\bar q}\cre{\bar s}\ann{r}\ann{p}\bigr) + \sum_{p\bar q\bar r\bar s}\bigl(g_{p\bar q\bar r\bar s}\cre{p}\cre{\bar r}\ann{\bar s}\ann{\bar q} + g_{p\bar q\bar r\bar s}\cre{\bar q}\cre{\bar s}\ann{\bar r}\ann{p}\bigr) 
\end{split}
\end{align}
$\hat H^\text{pb}_{\systa\systb}$ will be reduced based on size of the charge distributions, $\phi^*_P(\mathbf{x}_i)\phi_Q(\mathbf{x}_i)$, which enter the two-electron integrals (see eq \ref{eq:two_el_integral}). Integrals such as $g_{p\bar q r \bar s}$ will generally be much smaller than integrals such as $g_{p\bar q  \bar r \bar s}$ and $g_{p\bar q r  s}$. This is due to the fact that $g_{p\bar q r \bar s}$ contains an integral over two charge distributions, $\phi^*_p(\mathbf{x}_1)\phi_{\bar q}(\mathbf{x}_1)$ and $\phi^*_r(\mathbf{x}_2)\phi_{\bar s}(\mathbf{x}_2)$, which both are necessarily small since $p$ and $r$ are localized to the system, whereas $\bar q$ and $\bar s$ are localized to the environment. In contrast, $g_{p\bar q r  s}$  contains an integral over the  charge distributions, $\phi^*_p(\mathbf{x}_1)\phi_{\bar q}(\mathbf{x}_1)$ and $\phi^*_r(\mathbf{x}_2)\phi_{s}(\mathbf{x}_2)$, of which $\phi^*_r(\mathbf{x}_2)\phi_{s}(\mathbf{x}_2)$ will be large, in particular for $r=s$.  Retaining only the dominant terms, and wrapping it into an effective one-electron operator, we obtain a form of $\hat V$ which connects Fock space states which differ in the number of electrons by one. 
\begin{align}
\label{eq:interaction_approximated}
    \hat V = \sum_{p\bar q}(V_{p\bar q}\cre{p}\ann{\bar q} + V_{p\bar q}^*\cre{\bar q}\ann{p})
\end{align} 
The effective integral $V_{p\bar q}$ contains one-electron components as well as two-electron components of a charge distribution $\phi^*_p(\mathbf{x}_i)\phi_{\bar q}(\mathbf{x}_i)$ interacting with electrons in both the molecule and the environment. 
$\hat V$ has the bilinear form often found in descriptions of bipartite open quantum systems\cite{dzhioev2012nonequilibrium,elenewski2017communication,Philbin_Levy_Narang_Dou_2021,kilin2000electron,novoderezhkin2007mixing,dzhioev2011super,Pedersen_2022_redfield_propagation_photoinduced_ET}, but in this derivation, we have kept the physical content of it in terms of standard one- and two-electron integrals. However, note that although we know the content of  $V_{p\bar q}$, it can not be calculated without doing a full quantum mechanical calculation of the combined system and environment. It will therefore be parametrized based on knowledge of the magnitudes of the type of interactions it contains.

\subsection{Density operator formalism for FCI states}

We allow the two subsystems to interact by turning on the interaction $\hat V$ at some time $t=0$. We will now construct the density operator for the full system from the eigenstates of $\hat H_\systa$ and $\hat H_\systb$.
Using the effective interaction operator given in eq \ref{eq:interaction_approximated}, the Hamiltonian for the interacting system $\systa+\systb$ is given by
\begin{align}
    \hat H(t) = \hat H_\systa + \hat H_\systb + \hat V(t) \equiv  \hat H_0 + \hat V(t)
\end{align}
Turning on $\hat V(t)$ at time $t=0$ is enforced by using the Heaviside step function, $\hat V(t) = \theta(t)\hat V$. We regard $\hat H_0$ as the zeroth-order Hamiltonian and $\hat V(t)$ as a small perturbation. I.e., we assume that the interaction is weak, which is valid for the considered non-covalent interactions. We may construct common eigenstates for $\hat H_0$ from the eigenstates of $\hat H_\systa$ and $\hat H_\systb$ since $\hat H_0$ is additively separable.
The eigenstates for $\hat H_\systa$ and $\hat H_\systb$ are given by the FCI solutions
\begin{align}
    \mathbf H_\systa\mathbf C^\systa_a &= E_a^\systa \mathbf C^\systa_a\\
    \mathbf H_\systb\mathbf C^\systb_{\bar a} &= E_{\bar a}^\systb \mathbf C^\systb_{\bar a}
\end{align}
where the elements of the Hamiltonians are $\bra{I}\hat H_\systa\ket{J}$ and $\bra{\bar I}\hat H_\systb\ket{\bar J}$. The determinants are defined by $\ket{I}=\hat{A}_I^\dagger\vacket$ and $\ket{\bar I} = \hat{A}_{\bar I}^\dagger\vacket$, where $\hat{A}_I^\dagger$ and $\hat{A}_{\bar I}^\dagger$ are strings of creation operators for the system and environment, respectively. 
We thus have states for the system, $\ket{a}$, and the environment, $\ket{\bar a}$,
\begin{align}
    \ket{a} &= \sum_IC_{Ia}^\systa\ket{I}\label{eq:system_a_eigenstates}\\
    \ket{\bar a} &= \sum_{\bar I}C_{\bar I\bar a}^\systb\ket{\bar I}\label{eq:system_b_eigenstates}
\end{align}
which satisfy $\braket{a}{b}=\delta_{ab}$ and $\braket{\bar a}{\bar b}=\delta_{\bar a\bar b}$. Using these solutions, we construct a combined basis for the system and environment
\begin{align}
    \ket{a\bar a} = \sum_{I\bar I}C_{Ia}^\systa C_{\bar I\bar a}^\systb\hat{A}_{I}^\dagger\hat{A}_{\bar I}^\dagger\vacket
\end{align}
The orthonormality $\braket{a\bar a}{b\bar b}=\delta_{ab}\delta_{\bar a\bar b}$ follows from the properties of the FCI states (eqs \ref{eq:system_a_eigenstates} - \ref{eq:system_b_eigenstates}).
We note that the states $\ket{a\bar a}$ are not eigenstates of the full Hamiltonian $\hat H$, but they are eigenstates of $\hat H_0$, 
\begin{align}
    \hat H_0\ket{a\bar a} = \bigl(E_{a}^\systa + E_{\bar a}^\systb\bigr)\ket{a\bar a}
\end{align}
We build the density operator for the combined system from these zeroth-order solutions. In this basis, the density operator can be written as
\begin{align}
    \rho(t) = \sum_{a\bar a b\bar b}\rho_{a\bar a,b\bar b}(t)\ket{a\bar a}\bra{b\bar b}
\end{align}
This form of the density operator will be used to construct a reduced density operator for the system, implicitly taking the environment into account.

\subsection{The Liouville-von Neumann equation for the density operator}
\label{sec:LvN_for_density_operator}

The time evolution of the density operator is governed by the Liouville-von Neumann equation, which in the interaction picture is given by
\begin{align}
\label{eq:lvn_interaction_picture}
    \ddt\tilde\rho(t) = -i\comm{\tilde V(t)}{\tilde\rho(t)}
\end{align}
An operator $\hat \Omega$ in the interaction picture is given by the transformation $\tilde\Omega(t)=e^{i\hat H_0t}\hat\Omega e^{-i\hat H_0t}$. The goal is to obtain an equation for the time evolution of the molecular system $\systa$. To do this, we use the projection operator technique of Nakajima and Zwanzig\cite{nakajima1958quantum,Zwanzig_1960_projection_operators}, where we define two orthogonal projection superoperators $\pproj$ and $\qproj$. 
$\pproj$ is explicitly constructed such that the $\pproj$-projected density operator describes the dynamics of the molecule under the influence of the environment. This is achieved by
\begin{align}
\begin{split}
\label{eq:p_projection_operator}
    \pproj\tilde\rho(t) &\equiv \sum_{a\bar a b}\ket{a\bar a}\bra{a\bar a}\tilde\rho(t)\ket{b\bar a}\bra{b\bar a}\\
    &= \sum_{a\bar a b}\rho_{a\bar a,b\bar a}(t)e^{i\omega_{ab}t}\ket{a\bar a}\bra{b\bar a}
\end{split}
\end{align}
The exponential $e^{i\omega_{ab}t}$ results from the interaction picture, where $\omega_{ab}$ is the energy difference of the unperturbed molecule eigenstates, $\omega_{ab} = E_{a}^\systa-E_{b}^\systa$. From eq \ref{eq:p_projection_operator}, we see that the $\pproj$ superoperator effectively averages out the environmental degrees of freedom. $\qproj$ is defined as the orthogonal complement of $\pproj$,
\begin{align}
    \qproj\tilde\rho(t) &= (1-\pproj)\tilde\rho(t)
\end{align}
Since we only will consider weakly interacting subsystems, we approximate the elements $\rho_{a\bar a,b\bar a}(t)$ in eq \ref{eq:p_projection_operator} as a product
\begin{align}
\label{eq:density_operator_decomposition}
    \rho_{a\bar a,b\bar a}(t) \approx \sigma_{ab}(t)\rho_{\bar a \bar a}^\systb
\end{align}
where $\sigma_{ab}(t)$ are the elements of the reduced density matrix for the molecule. It is also assumed that the environment is large and has a short relaxation time relative to the system, such that the environment is unchanged by the interaction. Thus, the elements $\rho_{\bar a\bar a}^\systb$ are time-independent. 
To get a reduced description of the molecule, a decomposition of the density operator $\rho(t)$ is essential.\cite{breuer_2002_theory_of_open_quantum_systems,nitzan_2013_chemical_dynamics_in_condensed_phases}

We can construct two coupled equations by projecting the Liouville-von Neumann equation (eq \ref{eq:lvn_interaction_picture}) with $\pproj$ and $\qproj$,
\begin{align}
    \ddt\pproj\tilde\rho(t) &= -i\pproj\comm{\tilde V(t)}{\pproj\tilde\rho(t)} - i\pproj \comm{\tilde V(t)}{\qproj\tilde\rho(t)}\label{eq:p_projected_liouville}\\
    \ddt\qproj\tilde\rho(t) &= -i\qproj\comm{\tilde V(t)}{\pproj\tilde\rho(t)} - i\qproj \comm{\tilde V(t)}{\qproj\tilde\rho(t)}\label{eq:q_projected_liouville}
\end{align}
where we also have inserted the identity $1=\pproj+\qproj$ in front of the $\tilde\rho(t)$ inside the commutator. Combined, these equations are still equivalent to the Liouville-von Neumann equation for $\tilde\rho(t)$ and are as challenging to solve as the full problem. However, they serve as a convenient starting point for constructing an equation for the elements $\sigma_{ab}(t)$ of the reduced density matrix for the molecule.

\subsection{The Redfield equation for the reduced density matrix}
\label{sec:redfield_equation}

We will now introduce approximations into eqs \ref{eq:p_projected_liouville} - \ref{eq:q_projected_liouville}. A formal integration of eq \ref{eq:q_projected_liouville} gives
\begin{align}
\begin{split}
\label{eq:q_projected_liouville_integrated}
    \qproj\tilde\rho(t) = -i\int_0^t\qproj\comm{\tilde V(t')}{\pproj\tilde\rho(t')}dt'
    -i\int_0^t\qproj\comm{\tilde V(t')}{\qproj\tilde\rho(t')}dt'
\end{split}
\end{align}
where $\qproj\tilde\rho(0)=0$ since the interaction $V(t)$ is turned on at time $t=0$. Following the steps in the derivation of the Redfield equation\cite{nitzan_2013_chemical_dynamics_in_condensed_phases}, we assume the weak coupling limit, i.e., the interaction $\hat V$ is small compared to the energies of the molecule $E_a^\systa$. We approximate the above equation to the lowest order in $\hat V$ by disregarding the last integral, such that $\qproj\tilde\rho(t)$ no longer depends on itself. We further assume that the environment is large and unchanged by the interaction $\hat V$.
Inserting eq \ref{eq:q_projected_liouville_integrated} (excluding the omitted integral) back into eq \ref{eq:p_projected_liouville}, we obtain
\begin{align}
\label{eq:redfield_equation_double_commutator}
    \ddt\pproj\tilde\rho(t) = -\int_0^t\pproj\comm{\tilde V(t)}{\comm{\tilde V(t')}{\pproj\tilde\rho(t')}}dt'
\end{align}
where we have used that $\pproj \tilde V(t) \pproj\tilde\rho(t)=0$. By further assuming that memory effects are negligible, we get in the Markovian limit the following equation for $\sigma_{ab}(t)$,
\begin{align}
\label{eq:redifeld_equation_general}
    \ddt\sigma_{ab}(t) = -i\omega_{ab}\sigma_{ab}(t) +\sum_{cd}R_{abcd}\sigma_{cd}(t)
\end{align}
$\mathbf{R}$ is the Redfield tensor defined through the tensors $\boldsymbol\Gamma^+$ and $\boldsymbol\Gamma^-$,
\begin{align}
\begin{split}
    R_{abcd} = &(-1)^{\Delta N_{cd}}\bigl(\Gamma_{dbac}^+(\omega_{da}) + \Gamma_{dbac}^-(\omega_{bc})\\
        &- 
        \delta_{bd}\sum_e\Gamma_{aeec}^+(\omega_{de})-\delta_{ac}\sum_e\Gamma_{deeb}^-(\omega_{ec})
    \bigr)
\end{split}
\end{align}
$\Delta N_{cd}$ is the difference in the numbers of electrons between states $\ket{c}$ and $\ket{d}$. The $\boldsymbol\Gamma^x$ tensors ($x=+,-$) are defined by
\begin{align}
\begin{split}
    \Gamma_{abcd}^x(\omega) = \sum_{pq}
        &\bra{a}\cre{p}\ket{b}\bra{c}\ann{q}\ket{d}\int_0^\infty  e^{i\omega\tau}C_{pq}(x\tau)d\tau\\
        + 
        \sum_{pq}&\bra{a}\ann{q}\ket{b}\bra{c}\cre{p}\ket{d}\int_0^\infty  e^{i\omega\tau}D_{pq}(x\tau)d\tau
\end{split}
\end{align}
By introducing the notation $\langle\cdot \rangle_\systb$ for the environment average (see Supporting Information for details), we define the time-correlation functions $C_{pq}(\tau)$ and $D_{pq}(\tau)$,
\begin{align}
    C_{pq}(\tau) &= \sum_{\bar p\bar q}V_{p\bar p}V_{q\bar q}^*\bigl\langle\tann{\bar p}(\tau)\tcre{\bar q}(0)\bigr\rangle_\systb\label{eq:C_correlation_function}\\
    D_{pq}(\tau) &= \sum_{\bar p\bar q}V_{p\bar p}V_{q\bar q}^*\bigl\langle\tcre{\bar q}(\tau)\tann{\bar p}(0)\bigr\rangle_\systb\label{eq:D_correlation_function}
\end{align}

Since we are interested in steady-state properties (long time scales) of the reduced density matrix, we continue by assuming the secular approximation in eq \ref{eq:redifeld_equation_general}. This guarantees a master equation of Lindblad form, thereby overcoming any potential positivity issues of the Redfield equation.\cite{hartmann2020accuracy} The secular approximation exploits that off-diagonal elements of the reduced density matrix either oscillate rapidly compared to the evolution of the diagonal elements or decay to zero. Hence, the off-diagonal elements may be disregarded.\cite{jeske2012dual}
Eq \ref{eq:redifeld_equation_general} then simplifies to
\begin{align}
\label{eq:redifeld_equation_secular}
    \ddt\sigma_{aa}(t) &= \sum_b k_{a\leftarrow b}\sigma_{bb}(t)
    - \sum_b k_{b\leftarrow a}\sigma_{aa}(t)
\end{align}
where the transfer rates are defined by (see Supporting Information)
\begin{align}
\begin{split}
\label{eq:population_transfer_rate}
    k_{b\leftarrow a}
    = \sum_{pq}
        &\bra{a}\cre{p}\ket{b}\bra{b}\ann{q}\ket{a}\int_{-\infty}^\infty  e^{i\omega_{ab}\tau}C_{pq}(\tau)d\tau\\
        + 
        \sum_{pq}&\bra{a}\ann{q}\ket{b}\bra{b}\cre{p}\ket{a}\int_{-\infty}^\infty  e^{i\omega_{ab}\tau}D_{pq}(\tau)d\tau
\end{split}
\end{align}
Eq \ref{eq:redifeld_equation_secular} now has the form of a Pauli master equation.\cite{pauli1928mastereq,breuer_2002_theory_of_open_quantum_systems} By collecting the diagonal elements (populations) in a vector $\mathbf{P}$ with elements $P_a=\sigma_{aa}$, and the transfer rates in a matrix $\mathbf{W}$ with off-diagonal elements $W_{ab}=k_{a\leftarrow b}$ and diagonal elements $W_{aa}=-\sum_bk_{b\leftarrow a}$, we can write the equation as
\begin{align}
\label{eq:pauli_master_equation}
    \ddt \mathbf{P} = \mathbf{WP}
\end{align}
Due to the form of the diagonal elements $W_{aa}$, the matrix $\mathbf{W}$ will always contain linearly dependent rows. This entails that at least one steady-state solution can be found by solving for its null space.

\subsection{A mean-field approximation to the Redfield equation}
\label{sec:mean_field_approximation}

The above derivation assumes that we have the FCI states for the non-interacting subsystems, and the introduced approximations so far have been those of the Redfield equation. We will now start by making approximations for the environment by assuming mean-field states and a continuous spectrum. The correlation functions in eqs \ref{eq:C_correlation_function} - \ref{eq:D_correlation_function} can be simplified to (see Supporting Information)
\begin{align}
    \label{eq:correlation_function_minus_simplified}
    C_{pq}(\tau) &= \int d\omega V_{p}(\omega)V_{q}^*(\omega)e^{-i\omega\tau}[1-\bar n(\omega)]\\
    \label{eq:correlation_function_plus_simplified}
    D_{pq}(\tau) &= \int d\omega V_p(\omega)V_q^*(\omega)e^{i\omega\tau}\bar n(\omega)
\end{align}
where $\bar n(\omega)$ is the average number of electrons at the given energy $\omega$, and $V_p(\omega)$ is the continuous representation of $V_{p\bar p}$. For a given finite temperature $T$ and a chemical potential $\mu$, the average number of electrons can, e.g., be given by the Fermi-Dirac distribution,
\begin{align}
    \bar n(\omega) = \Bigl[\exp\Bigl(\frac{\omega-\mu}{k_BT}\Bigr)+1\Bigr]^{-1}
\end{align}
with $k_B$ being the Boltzmann constant. By inserting the simplified correlation functions in eqs \ref{eq:correlation_function_minus_simplified} - \ref{eq:correlation_function_plus_simplified} into the expression for the population transfer rates in eq \ref{eq:population_transfer_rate} we obtain
\begin{align}
\begin{split}
\label{eq:population_transfer_rate_2} 
    k_{b\leftarrow a}
    = \sum_{pq}
        &
        \bra{a}\cre{p}\ket{b}\bra{b}\ann{q}\ket{a} J_{pq}(\omega_{ab})[1-\bar n(\omega_{ab})]\\
        +\sum_{pq}&
        \bra{a}\ann{q}\ket{b}\bra{b}\cre{p}\ket{a} J_{pq}(\omega_{ba})\bar n(\omega_{ba})
\end{split}
\end{align}
where we have introduced $J_{pq}(\omega) \equiv 2\pi V_p(\omega)V_q^*(\omega)$, which we will refer to as the spectral density function.

We now approximate the electronic states of the molecule by using single determinants expressed in the same spin orbital basis. The transfer rates in eq \ref{eq:population_transfer_rate_2} can then be simplified by noting that $\bra{a}\cre{p}\ket{b}\bra{b}\ann{q}\ket{a}=0$ if $p\neq q$. Since the states $\ket{a}$ and $\ket{b}$ are expressed in the same spin orbital basis, the energy difference is $\omega_{ab}=\varepsilon_p$, where $\varepsilon_p$ is the orbital energy of $\phi_p$. 
By writing $J_p(\omega) \equiv J_{pp}(\omega) = 2\pi|V_p(\omega)|^2$, the resulting expression for the population transfer rates is
\begin{align}
\label{eq:transfer_rate_single_determinants}
\begin{split}
    k_{b\leftarrow a}
    = \sum_{p}\Bigl(
        &
        \bra{a}\cre{p}\ket{b}\bra{b}\ann{p}\ket{a} J_{p}(\eps_p)[1-\bar n(\eps_p)]\\
        +&
        \bra{a}\ann{p}\ket{b}\bra{b}\cre{p}\ket{a} J_{p}(\eps_p)\bar n(\eps_p)
    \Bigr)
\end{split}
\end{align}
In principle, one may include all determinants. However, the choice of which determinants to include may be motivated by the fact that high-energy (relative to the chemical potential of the environment) spin orbitals will remain unoccupied. Similarly, low-energy spin orbitals will remain occupied. Spin-orbitals with energies close to the chemical potential therefore constitute an active space for constructing relevant determinants for the reduced density matrix.

\subsubsection{The spectral density function}
\label{sec:spectral_density_function}

The spectral density function $J_p(\omega)$ contains information about the environment's density of states and the interaction strength of the system orbitals with the environment. 
In this work, we approximate the interaction strengths to be equal for all spin orbitals. In this case, the spectral density function becomes proportional to the environment's electronic density of states. The density of states can either be modeled or measured experimentally.\cite{berglund1964photoemission,reinert2005photoemission} We will estimate the density of states for the considered molecular environment by calculating its relative number of orbitals per some fixed energy spacing. In the wide band approximation, $J_p(\omega)$ is set to a constant. This is a reasonable choice for metallic environments such as gold or silver with a near-constant density of states around the Fermi energy.\cite{verzijl2013applicability,smith1974photoemission}

\subsubsection{Energy level broadening}

Excited states have finite lifetimes, and therefore broadened energy levels, due to interactions not captured by the electronic Hamiltonian.\cite{rotter2009non,nordlander1990energy} In electronic structure theory, this has been accounted for by introducing phenomenological damping.\cite{orr1971perturbation,norman2005nonlinear} This relaxes the resonance requirement for transitions between molecule states. 
We incorporate this effect by replacing Dirac delta functions in the time-correlation functions with normalized Gaussians with standard deviations $\gamma$ (see Supporting Information). Eq \ref{eq:transfer_rate_single_determinants} is therefore modified to
\begin{align}
\label{eq:transition_rate_with_energy_level_broadening}
\begin{split}
    k_{b\leftarrow a}
    = \frac{1}{\sqrt{2\pi\gamma^2}}\sum_{p}\Bigl(&\bra{a}\cre{p}\ket{b}\bra{b}\ann{p}\ket{a}\int_{-\infty}^\infty J_{p}(\omega)\exp\Bigl(-\frac{(\omega_{ab}-\omega)^2}{2\gamma^2}\Bigr)[1-\bar n(\omega)]d\omega\\
        +
        &\bra{a}\ann{p}\ket{b}\bra{b}\cre{p}\ket{a}\int_{-\infty}^\infty J_{p}(\omega)\exp\Bigl(-\frac{(\omega_{ba}-\omega)^2}{2\gamma^2}\Bigr)\bar n(\omega)d\omega
    \Bigr)
\end{split}
\end{align}
We have now arrived at the final expression for the population transfer rates.

\section{Results and discussion}
\label{sec:results_and_discussion}

In this section, we illustrate how the Redfield equation can be used to describe the fractional charging of a molecular system upon interaction with an environment. We note that we include only the novel particle-breaking interactions described in Section \ref{sec:dividing_system}. However, particle-conserving interactions must also be considered for realistic chemical applications. As a model system, we consider a benzene molecule physisorbed on a finite graphene surface. The benzene molecule is the target for which we create a reduced density matrix, while the graphene sheet is the environment. We will explore how the fractional charging of benzene depends on the chemical potential, temperature, and energy level broadening.
In Section \ref{sec:computational_details}, we describe necessary computational details, while results are presented in Section
\ref{sec:average_number_of_electrons}. Future prospects of using the proposed formalism are discussed in Section
\ref{sec:challenges_and_limitations}.

\subsection{Computational details}
\label{sec:computational_details}

\begin{figure}[H]
    \centering
    \includegraphics[scale=1.4]{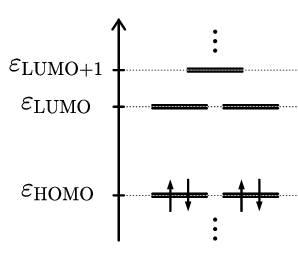}
    \caption{Energy level diagram for the two degenerate highest occupied molecular orbitals and the three lowest unoccupied molecular orbitals of the neutral benzene molecule. The illustrated electron configuration corresponds to the ground state. The orbital energies are, in atomic units, $\eps_\text{HOMO} = -0.336252$, $\eps_\text{LUMO} = 0.129394$, and $\eps_{\text{LUMO}+1} = 0.145115$.}
    \label{fig:MO_diagram}
\end{figure}

We will use a restricted Hartree-Fock calculation of isolated, neutral benzene (see Supporting Information for geometry specification) using the cc-pVTZ basis\cite{dunning1989gaussian} as a reference determinant. All determinants will be constructed from these spin orbitals. 
An active space of five molecular orbitals (i.e., ten spin orbitals) is chosen, see Figure \ref{fig:MO_diagram}. By distributing between zero and ten electrons among these orbitals, we obtain 32 possible determinants. The energy differences between the determinants are given by $\omega = \pm\varepsilon_p$, with $+$ for the addition and $-$ for the removal of an electron as mentioned in Section \ref{sec:mean_field_approximation}. 
 
\begin{figure}[H]
    \centering
    \includegraphics[scale=0.55]{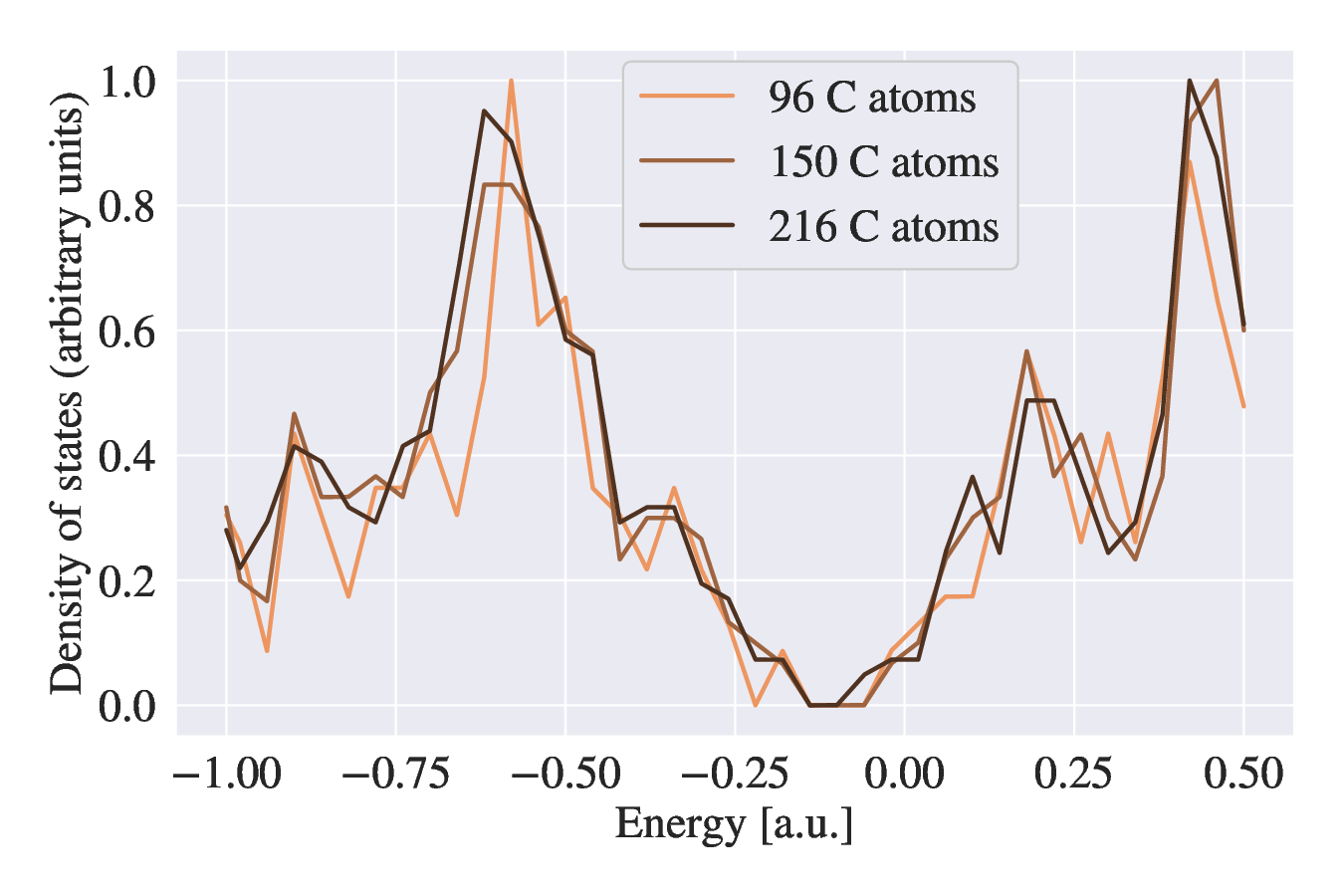}
    \caption{The approximated density of states calculated for graphene surfaces of different sizes. The density of states was calculated to be the relative number of orbitals within a window of 0.05 a.u. The orbitals were obtained from restricted Hartree-Fock calculations in the cc-pVDZ basis.}
    \label{fig:density of states_afo_n_c_atoms}
\end{figure}

We describe the graphene surface, which plays the role of the environment, in a simplified manner. We assume that the graphene is unaffected by the interaction with the benzene molecule, as motivated by the many degrees of freedom (see Section \ref{sec:LvN_for_density_operator}). Further, all spin orbitals are assumed to interact with the same strength. The spectral density function is estimated by a Hartree-Fock calculation on a finite graphene sheet using the cc-pVDZ basis\cite{dunning1989gaussian} and an energy spacing of 0.05 a.u. The density of states is calculated for sheets consisting of between 96 and 216 carbon atoms. Increasing the size of the graphene sheet does not change the density of states considerably, see Figure \ref{fig:density of states_afo_n_c_atoms}. We will therefore use the sheet containing 216 carbon atoms as the environment. The functional form resembles the reported density of states obtained from a density functional theory calculation presented in Ref. \citenum{liang2019oxygen}.

The Redfield equation (eq \ref{eq:pauli_master_equation}) using the population transfer rates of eq \ref{eq:transition_rate_with_energy_level_broadening} is propagated in time by Euler integration\cite{Griffiths2010} until steady state is reached. The initial condition for the time propagation is the neutral benzene molecule.


\subsection{Numerical illustration: Fractional charging of benzene at steady state}
\label{sec:average_number_of_electrons}

Here, we show how the presented model can induce fractional charging of a molecule. By using the steady-state solution of benzene, we illustrate how fractional charging depends on the chemical potential of the environment and energy level broadening. We note that realistic temperatures will not impact fractional charging. If the energy differences between the molecule's charge states are on the order of $\SI{1}{eV}$, inducing fractional charging would require temperatures around $\SI{e4}{\kelvin}$. At such elevated temperatures, our molecule would break down. Therefore, we set the temperature to a realistic constant, $T=\SI{300}{\kelvin}$, in our calculations.

\begin{figure}[H]
     \centering
     \begin{subfigure}[b]{\textwidth}
         \centering
         \includegraphics[scale=0.53]{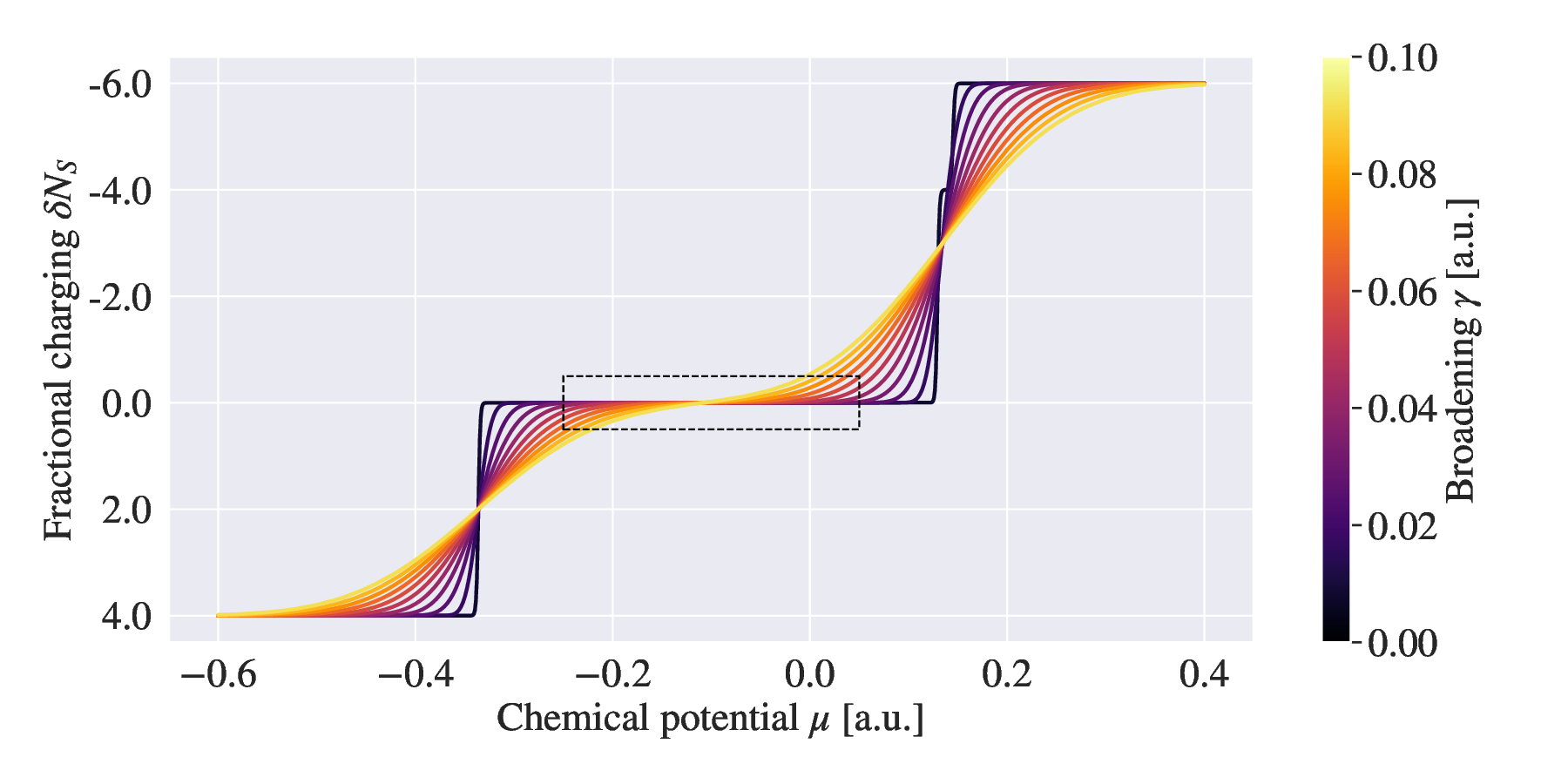}
         \caption{}
         \label{fig:dN_wba_large}
     \end{subfigure}
     \hfill
     \begin{subfigure}[b]{\textwidth}
         \centering
         \includegraphics[scale=0.53]{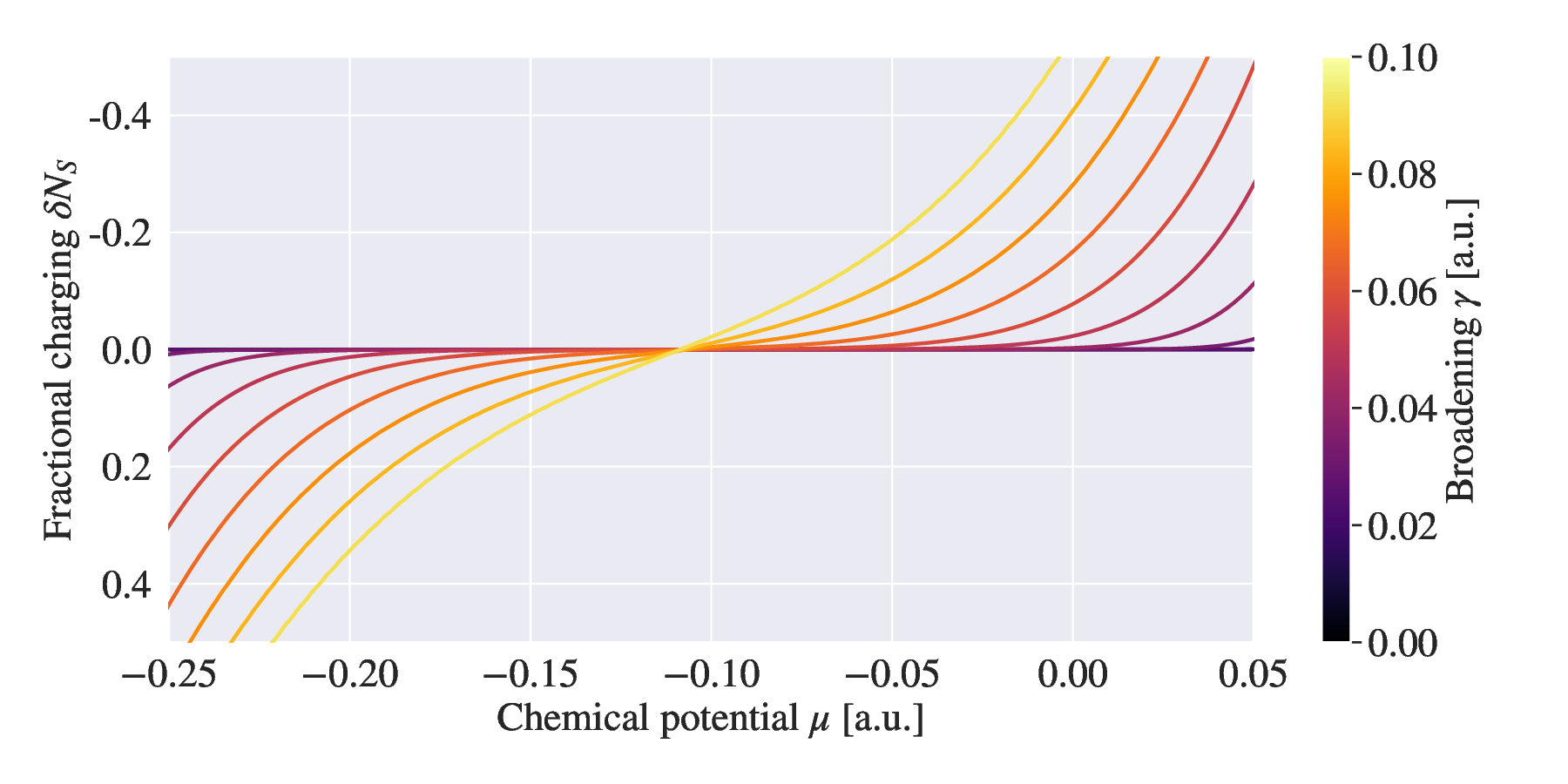}
         \caption{}
         \label{fig:dN_wba_small}
     \end{subfigure}
        \caption{Fractional charging as a function of the chemical potential of the environment at different energy-level broadenings $\gamma$ with the wide band approximation as the spectral density function. The region marked in a) is enlarged in b). The temperature was set to $T=\SI{300}{\kelvin}$.}
        \label{fig:dN_wba}
\end{figure}

First, we use the wide band approximation for the spectral density function, $J_p(\omega)$. The results are shown in Figure \ref{fig:dN_wba}. When the broadening $\gamma$ is small (close to zero) there is little to no fractional charging. In this case, the main effect is integer charging when the chemical potential crosses the energy differences of the integer charge states. When the chemical potential matches exactly the energy differences, the populations of the corresponding states are equal. 
The values for $\delta N_\systa$ at $\mu=0$ are listed in Table \ref{tab:fractional_charging}, where the dependence of $\delta N_\systa$ on the broadening is seen. Even at $\mu=0$, significant fractional charging effects are seen for $\gamma > 0.03$. By significant, we mean fractional charging effects that influence the energy on the order of $10^{-3}$ Hartree. 
We note that the fractional charging is zero when the chemical potential equals the midpoint between the HOMO and LUMO energy levels (see Figure \ref{fig:MO_diagram}), regardless of the broadening. The symmetry around this point is due to the symmetry of the wide band approximation. 

\begin{figure}[H]
     \centering
     \begin{subfigure}[b]{\textwidth}
         \centering
         \includegraphics[scale=0.53]{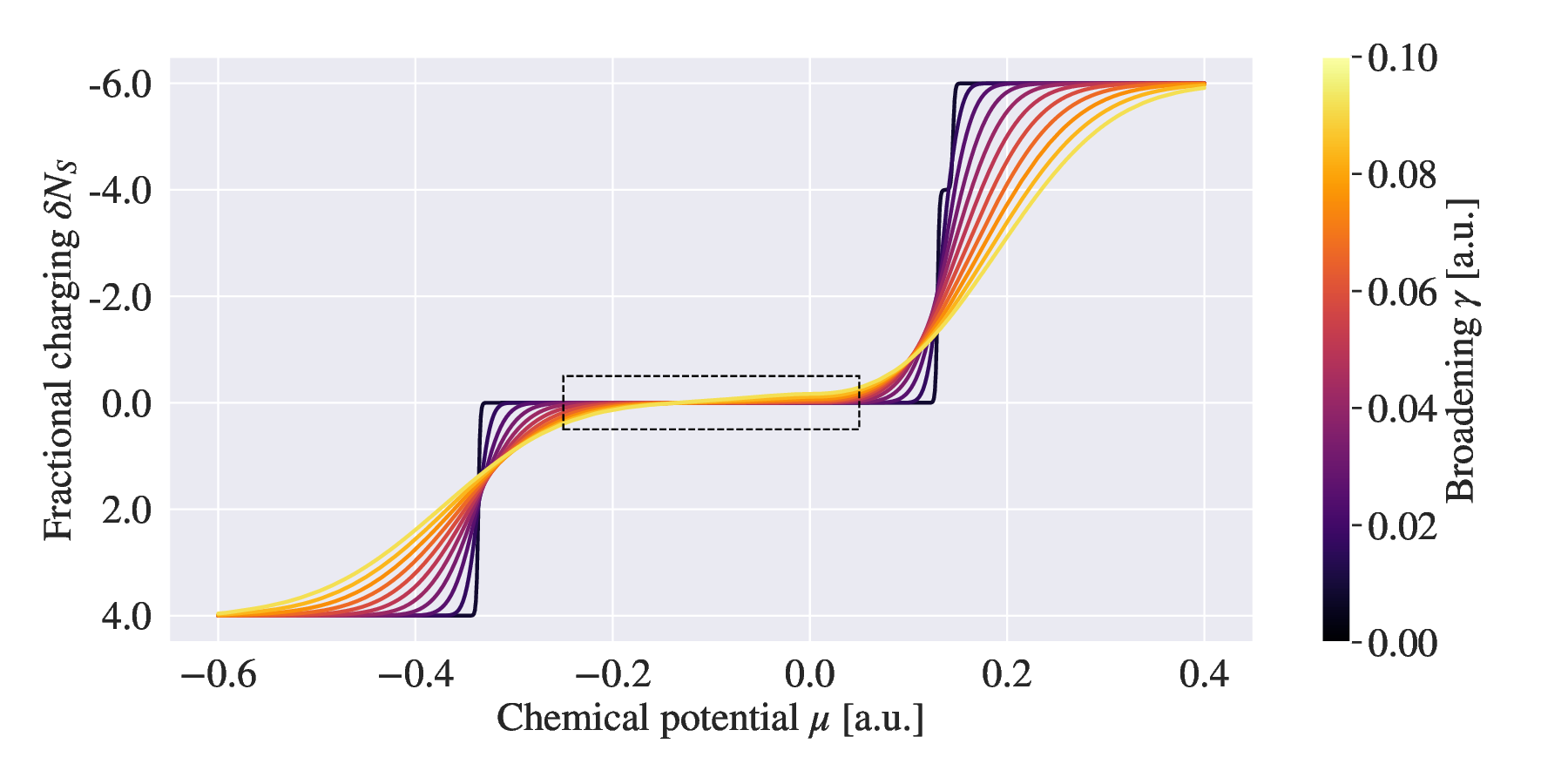}
         \caption{}
         \label{fig:dN_approximate_large}
     \end{subfigure}
     \hfill
     \begin{subfigure}[b]{\textwidth}
         \centering
         \includegraphics[scale=0.53]{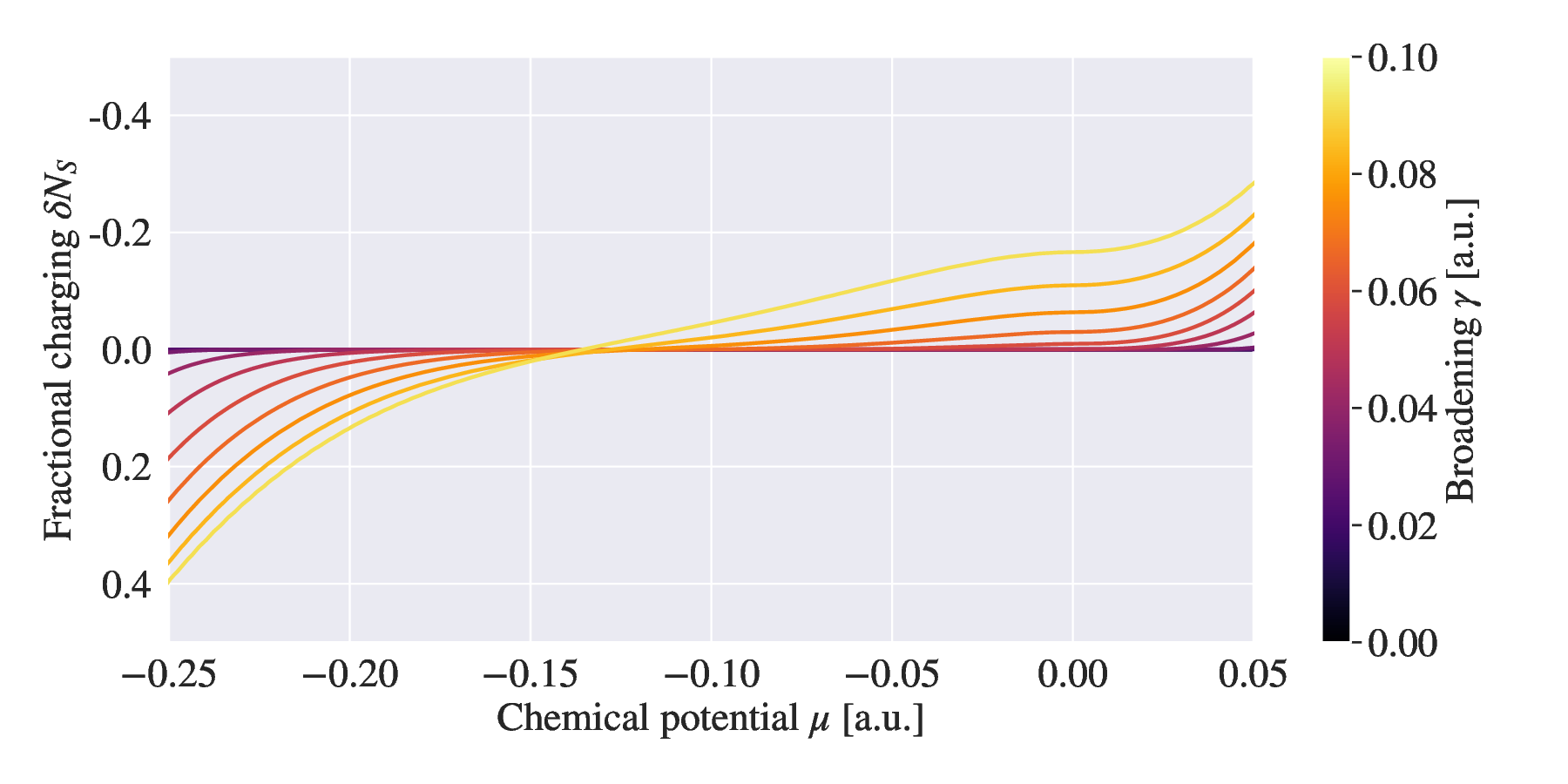}
         \caption{}
         \label{fig:dN_approximate_small}
     \end{subfigure}
        \caption{Fractional charging as a function of the chemical potential of the environment at different energy-level broadenings $\gamma$ with the calculated electronic density of states for graphene as the spectral density function. The region marked in a) is enlarged in b). The temperature was set to $T=\SI{300}{\kelvin}$.}
        \label{fig:dN_approximate}
\end{figure}

Figure \ref{fig:dN_approximate} shows the fractional charging as a function of the chemical potential where the calculated electronic density of states is used as the spectral density function. Using the density of states shows a qualitative difference compared to using the wide band approximation. Notably, at $\mu=0$, the same energy level broadening gives a less significant fractional charging (see Table \ref{tab:fractional_charging}). This is because the density of states (see Figure \ref{fig:density of states_afo_n_c_atoms}) is close to zero around $\mu=0$. Further, the curves are no longer symmetric about neither the energy differences between the charge states nor about the midpoint between the HOMO and LUMO energies. These asymmetries are a direct consequence of the asymmetry of the density of states. The qualitative differences show that the choice of the spectral density function and the broadening and how it is modeled significantly affects the properties of the system of interest. 

\begin{table}[H]
    \centering
    \begin{tabular}{c c c}
    \toprule
    Broadening $\gamma$ [a.u.] & $\delta N_\systa$ for WBA & $\delta N_\systa$ for DOS \\
    \midrule
    0.10	&      0.537   &   0.167 \\
    0.09	&      0.409   &   0.110 \\
    0.08	&      0.283   &   0.064 \\
    0.07	&      0.169   &   0.031 \\
    0.06	&      0.080   &   0.001 \\
    0.05	&      0.026   &   0.000 \\
    0.04	&      0.003   &   0.000 \\
    0.03	&      0.000   &   0.000 \\
    0.02	&      0.000   &   0.000 \\
    0.01	&      0.000   &   0.000 \\
    0.00	&      0.000   &   0.000 \\
    \bottomrule
    \end{tabular}
    \caption{Fractional charging $\delta N_\systa$ of benzene at zero chemical potential for different values of the energy level broadening $\gamma$. The results are shown for the two models for the spectral density function; the wide band approximation (WBA) and the calculated electronic density of state for graphene (DOS).}
    \label{tab:fractional_charging}
\end{table}

\subsection{Future prospects of using the Redfield equation for electronically open molecular systems}
\label{sec:challenges_and_limitations}

The presented results are based on several key approximations, which we will now address. These include the assumptions of the interaction operator, the system states, and the inherent assumptions of the Redfield equation. The approximations will be discussed in the context of describing a molecular system interacting weakly with an environment, and with the electronic structure of the molecule as the main objective. First, particle-conserving interactions (see Section \ref{sec:dividing_system}) must also be included to appropriately account for the interaction with the environment. The particle-breaking interactions allow for describing fractional charging, but as discussed previously, we expect them to be only a small correction to the total interaction. The effect of particle-conserving interactions can, for example, be included by well-established quantum mechanical embedding methods or by a classical atomistic approach\cite{hofener2012wfinwf,saether2017mlhf,myhre2014mlcc,jacob2014subsystem,lee2019projection,manby2012simple,senn2009qm,cappelli2016integrated,tzeliou2022qmmmreview}. Second, we have used single determinants with a common set of spin orbitals as the basis for the system's reduced density matrix. Knowing that single-determinantal descriptions lack electron correlation, using correlated wave function models in the Redfield equation must be explored. A correlated model (e.g., the configuration interaction or coupled cluster wave functions) would change the Redfield equation's dynamics and result in a different steady-state solution. Third, the optimal spin orbitals change when we get a mixture of multiple charge states. However, relaxing the spin orbitals while propagating the Redfield equation will greatly increase the complexity and computational cost of the method. In addition, recomputing spin orbitals for the fractionally charged molecule would require a wave function method that handles non-integral electron numbers and fractional occupations. However, we expect the effect of fractional charging to be small such that the spin orbitals of the neutral molecule dominate.

The approximations for the Redfield equation may also be explored to assess its validity for describing fractional charging. The Born approximation is appropriate for non-covalent interactions between the system and a large environment that is unchanged by the interaction. Consequently, the Redfield equation is less suitable for describing environments consisting of finite-sized molecules. For smaller environments, memory effects become more significant. Further, if the system and the environment are similar in size, the Markov approximation breaks down due to the inability to separate the time scales.\cite{breuer2016colloquium,de2017dynamics,zhang2019exact} Examining the validity of the Born and Markov approximations would require comparisons to more sophisticated methods, such as the hierarchical equations of motion.\cite{tanimura_kubo1989HEOM,zheng2012heom_rev}

Another simplification is the assumption of equal interaction strengths for all benzene orbitals. In reality, interactions vary with distance and energy, and molecular orbitals are not uniquely defined. However, relative interaction strengths must be modeled in some way since explicit treatment of the environment is undesired. In principle, calculation of electronic coupling elements can be achieved by methods such as constrained DFT\cite{wu2006extracting,kaduk2012constrained} or generalized Mulliken-Hush.\cite{mulliken1952molecular,hush1961adiabatic,hush1968homogeneous,cave1996generalization}

\section{Summary and conclusions}
\label{sec:conclusions}

In this work, we introduce a reduced density operator formalism to include the effects of fractional charging of a molecule due to a non-covalent interaction with its environment.  In order to develop the formalism, we start by analyzing the full configuration interaction solution in a spin orbital basis which is localized to either the molecule or the environment. In this basis, the terms of the molecular electronic Hamiltonian for the full system may be categorized into a molecule Hamiltonian, an environment Hamiltonian, and an interaction Hamiltonian. The interaction Hamiltonian may further be divided into particle-conserving and particle non-conserving (particle-breaking) terms by considering the commutators with the particle number operator for the molecule. The particle-breaking interactions and the fractional charging they induce are the focus of this work, and the paper thus addresses a type of interaction that has largely been neglected by the quantum chemistry community.

We include the particle-breaking interactions as a perturbation to the non-interacting molecule and environment. By introducing an explicit form of the projection superoperator  we may use the Nakajima-Zwanzig formalism from the theory of open quantum systems. This allows us to derive an equation for the reduced density operator of the molecule starting from the Liouville-von Neumann equation for the full density operator. We employ the Born and Markov approximations to obtain a Redfield equation for the populations of the reduced density operator,  where the effect of the environment is included approximately. The effect of the environment is modeled through the spectral density function. In this work, it is taken to be either a constant (the wide band approximation) or the computed electronic density of states for the environment. The fractional charging results from non-zero populations for molecular states with different integer charges and signifies that electrons are, to a certain extent, shared with the environment.  Further, a phenomenological broadening of energy levels is introduced to mimic the finite lifetimes of electronic states. We present numerical illustrations for a mean-field approach of the formalism, using benzene physisorbed on graphene as a toy model.

The work presented in this paper represents an important step toward understanding molecules as electronically open quantum systems. To achieve quantitative accuracy, it will in the future be necessary to include particle-conserving interactions and use correlated electronic states as the many-electron basis. 

\section*{Acknowledgements}
We thank Associate Professor Thorsten Hansen for fruitful discussions regarding Redfield theory. 
I-M.H and B.S.S. acknowledge funding from the Research Council of Norway through \\FRINATEK project 325574. I-M.H acknowledges support from the Centre for Advanced Study in Oslo, Norway, which funded and hosted her Young CAS Fellow research project during the academic year of 22/23 and 23/24. J.P. acknowledges financial support from the Technical University of Denmark within the Ph.D. Alliance Programme, Knud Højgaards Fond (grant no. 23-02-2571), William Demant Fonden (grant no. 23-3137), and Ingeniør Alexandre Haynman og fru Nina Haynmans Fond (grant no. KR-4110605).

\bibliography{main.bib}

\appendix

\input{SI.tex}

\end{document}

%% file: SI.tex
\section*{Supporting Information --\\
Deriving the Redfield Equation for Electronically Open Molecules}

\section{The interaction Hamiltonian}
\label{sec:SI_interaction_Hamiltonian}

We show how we can approximate the interaction Hamiltonian between the molecular system $\systa$ and the environment $\systb$ as a bilinear operator. The full Hamiltonian is given by
\begin{align}
    \hat H = \sum_{PQ}h_{PQ}\cre{P}\ann{Q} + \half\sum_{PQRS}g_{PQRS}\cre{P}\cre{R}\ann{S}\ann{Q}
\end{align}
Assuming localized orbitals assigned to either the molecule (denoted with lowercase unbarred indices) or the environment (unbarred indices), the Hamiltonian can be separated as
\begin{align}
    \hat H = \hat H_\systa + \hat H_\systb + H_{\systa\systb}
\end{align}
where 
\begin{align}
    \hat H_\systa &= \sum_{pq}h_{pq}E_{pq} + \half\sum_{pqrs}g_{pqrs}e_{pqrs}\\
    \hat H_\systb &= \sum_{\bar p \bar q}h_{\bar p\bar q}E_{\bar p\bar q} + \half\sum_{\bar p\bar q\bar r\bar s}g_{\bar p\bar q\bar r\bar s}e_{\bar p\bar q\bar r\bar s}\\
    \hat H_{\systa\systb} &= \hat H_{\systa\systb}^\text{pc} + \hat H_{\systa\systb}^\text{pb}
\end{align}
For convenience, the short-hand notation for the one- and two-electron excitation operators $E_{PQ}=\cre{P}\ann{Q}$ and $e_{PQRS}=\cre{P}\cre{R}\ann{Q}\ann{S}$ is introduced. $\hat H_{\systa\systb}$ is the interaction operator with the terms
\begin{align}
\begin{split}
    \hat H_{\systa\systb} = &\sum_{\bar pq}h_{\bar pq}E_{\bar pq} + \sum_{p\bar q}h_{p\bar q}E_{p\bar q}
    + \half\sum_{\bar pqrs}g_{\bar pqrs}e_{\bar pqrs} + \half\sum_{p\bar qrs}g_{p\bar qrs}e_{p\bar qrs}\\
    + \half&\sum_{pq\bar rs}g_{pq\bar rs}e_{pq\bar rs} + \half\sum_{pqr\bar s}g_{pqr\bar s}e_{pqr\bar s}
    + \half\sum_{\bar p\bar qrs}g_{\bar p\bar qrs}e_{\bar p\bar qrs} + \half\sum_{\bar pq\bar r s}g_{\bar pq\bar r s}e_{\bar pq\bar r s}\\
    + \half&\sum_{\bar pqr\bar s}g_{\bar pqr\bar s}e_{\bar pqr\bar s} + \half\sum_{p\bar q\bar rs}g_{p\bar q\bar rs}e_{p\bar q\bar rs}
    + \half\sum_{p\bar qr\bar s}g_{p\bar qr\bar s}e_{p\bar qr\bar s} + \half\sum_{pq\bar r\bar s}g_{pq\bar r\bar s}e_{pq\bar r\bar s}\\
    + \half&\sum_{\bar p\bar q\bar rs}g_{\bar p\bar q\bar rs}e_{\bar p\bar q\bar rs} + \half\sum_{\bar p\bar qr\bar s}g_{\bar p\bar qr\bar s}e_{\bar p\bar qr\bar s}
    + \half\sum_{\bar pq\bar r\bar s}g_{\bar pq\bar r\bar s}e_{\bar pq\bar r\bar s} + \half\sum_{p\bar q\bar r\bar s}g_{p\bar q\bar r\bar s}e_{p\bar q\bar r\bar s}.
\end{split}
\end{align}
We decompose the full interaction operator into its particle-conserving and particle-breaking components. The particle-conserving terms as
\begin{align}
\begin{split}
    \hat H_{\systa\systb}^\text{pc} = 
    \half\sum_{pq\bar r\bar s}(g_{pq\bar r\bar s}e_{pq\bar r\bar s} + g_{qp\bar s\bar r}e_{qp\bar s\bar r}) + \half\sum_{p\bar q\bar rs}(g_{p\bar q\bar rs}e_{p\bar q\bar rs} + g_{\bar qps\bar r}e_{\bar qps\bar r})
\end{split}
\end{align}
and the particle-breaking terms as
\begin{align}
\begin{split}
    \hat H_{\systa\systb}^\text{pb} =&\sum_{p\bar q}(h_{p\bar q}E_{p\bar q} + h_{\bar qp}E_{\bar qp})
    + \sum_{p\bar qrs}(g_{p\bar qrs}e_{p\bar qrs} + g_{\bar qpsr}e_{\bar qpsr})\\
    + \half&\sum_{p\bar qr\bar s}(g_{p\bar qr\bar s}e_{p\bar qr\bar s} + g_{\bar qp\bar sr}e_{\bar qp\bar sr})
    + \sum_{p\bar q\bar r \bar s}(g_{p\bar q\bar r \bar s}e_{p\bar q\bar r \bar s} + g_{\bar qp\bar s \bar r}e_{\bar qp\bar s \bar r})
\end{split}
\end{align}
We will only focus on the particle-breaking interactions. First, we disregard the terms including $g_{p \bar qr \bar s}e_{p \bar qr \bar s}$ and $g_{\bar p q\bar r s}e_{\bar p q\bar r s}$ since the charge distributions of the integrals are separated in space and are thus expected to be small compared to the rest. Keeping the dominating terms of the interaction, we have
\begin{align}
\begin{split}
    \hat H_{\systa\systb}^\text{pb} = \sum_{p\bar q}(h_{p\bar q}E_{p\bar q} + h_{\bar qp}E_{\bar qp})
    + \sum_{p\bar qrs}(g_{p\bar qrs}e_{p\bar qrs} + g_{\bar qpsr}e_{\bar qpsr})
    + \sum_{p\bar q\bar r\bar s}(g_{p\bar q\bar r\bar s}e_{p\bar q\bar r\bar s} + g_{\bar qp\bar s\bar r}e_{\bar qp\bar s\bar r})
\end{split}
\end{align}
Using the anticommutation relations of the creation and annihilation operators and renaming the dummy summation indices, we can rewrite the expression as
\begin{align}
\begin{split}
    \hat H_{\systa\systb}^\text{pb} = &\sum_{p\bar q}\Bigl(
        h_{p\bar q} + \sum_{\bar r \bar s}g_{p\bar q\bar r\bar s}E_{\bar r\bar s} + \sum_{rs}g_{p\bar q rs}E_{rs} - \sum_{r}g_{prr\bar q}
    \Bigr)E_{p\bar q}\\
    +&\sum_{p\bar q}E_{\bar qp}\Bigl(
        h_{\bar qp} + \sum_{\bar r \bar s}g_{\bar qp\bar s\bar r}E_{\bar s\bar r} + \sum_{rs}g_{\bar qp sr}E_{sr} - \sum_{r}g_{rp\bar qr}
    \Bigr)
\end{split}
\end{align}
We  construct an effective interaction operator
\begin{align}
\begin{split}
    \hat H_{\systa\systb,\text{eff}}^\text{pb} &= \sum_{p\bar q}\Bigl(
        h_{p\bar q} + \sum_{\bar i}g_{p\bar q\bar i\bar i} + \sum_{i}g_{p\bar q ii} - \sum_{r}g_{prr\bar q}
    \Bigr)\cre{p}\ann{\bar q}\\
    &+ \sum_{p\bar q}\cre{\bar q}\ann{p}\Bigl(
        h_{\bar qp} + \sum_{\bar i}g_{\bar qp\bar i\bar i} + \sum_{i}g_{\bar qp ii} - \sum_{r}g_{rp\bar qr}
    \Bigr)
\end{split}
\end{align}
Finally, we define
\begin{align}
    V_{p\bar q} &= h_{p\bar q} + \sum_{\bar i}g_{p\bar q\bar i\bar i} + \sum_{i}g_{p\bar q ii} - \sum_{r}g_{prr\bar q}
\end{align}
to arrive at the bilinear form of the interaction operator
\begin{align}
    \hat V &= \sum_{p\bar q}(V_{\bar p q}\cre{\bar p}\ann{q} + V_{p\bar q}^*\cre{q}\ann{\bar p})
\end{align}

\section{Details on the derivation of the Redfield equation}
\label{sec:Redfield_derivation}

The differential equation for the time evolution of the projected density operator to second order in the interaction $\hat V$ is
\begin{align}
    \ddt\pproj\tilde\rho(t) = -\int_0^t\pproj\comm{\tilde V(t)}{\comm{\tilde V(t')}{\pproj\tilde\rho(t')}}dt'
\end{align}
We get an equation for the elements of the reduced density operator $\sigma$ by inserting our definitions of the projection superoperator $\pproj$ and the density operator $\tilde\rho(t)$ in the interaction picture. We project with the states $\bra{a\bar a}$ from the left and $\ket{b\bar a}$ from the right with a sum over all environment states to get
\begin{align}
\begin{split}
    &\ddt\sum_{\bar a}\bra{a\bar a}\pproj \bigl(e^{iH_0t}\rho(t)e^{-iH_0t}\bigr)\ket{b\bar a}\\
    =& e^{i\omega_{ab}t}Z_\systb\bigl(i\omega_{ab}\sigma_{ab}(t) + \ddt\sigma_{ab}(t)\bigl)
\end{split}
\end{align}
where $Z_\systb = \Tr{\systb}{\rho^\systb}$
Collecting terms, we get
\begin{align}
    \ddt\sigma_{ab}(t) = -i\omega_{ab}\sigma_{ab}(t) - \frac{e^{-i\omega_{ab}t}}{Z_\systb}\int_0^t\bra{a}\pproj\comm{\tilde V(t)}{\comm{\tilde V(t')}{\pproj\tilde\rho(t')}}\ket{b}dt'
\end{align}
The terms arising from the double commutator are
\begin{align}
\label{eq:double_commutator_written_out}
\begin{split}
    \frac{e^{-i\omega_{ab}t}}{Z_\systb}\bra{a}\pproj\comm{\tilde V(t)}{\comm{\tilde V(t')}{\pproj\tilde\rho(t')}}\ket{b}dt' = \frac{e^{-i\omega_{ab}t}}{Z_\systb}\sum_{\bar a}\Bigl(&\bra{a\bar a}\pproj\bigl(\tilde V(t)\tilde V(t')\pproj\tilde\rho(t')\bigr)\ket{b\bar a}\\
    -&\bra{a\bar a}\pproj\bigl(\tilde V(t)\pproj\tilde\rho(t')\tilde V(t')\bigr)\ket{b\bar a}\\
    -&\bra{a\bar a}\pproj\bigl(\tilde V(t')\pproj\tilde\rho(t')\tilde V(t)\bigr)\ket{b\bar a}\\
    +&\bra{a\bar a}\pproj\bigl(\pproj\tilde\rho(t')\tilde V(t')\tilde V(t)\bigr)\ket{b\bar a}\Bigr)
\end{split}
\end{align}
We start by looking at the first of these four terms. Inserting definitions for $\pproj$ and $\rho$, and using the fact that the states are eigenstates of $\hat H_0$,
\begin{align}
\begin{split}
    &\frac{e^{-i\omega_{ab}t}}{Z_\systb}\sum_{\bar a}\bra{a\bar a}\pproj\bigl(\tilde V(t)\tilde V(t')\pproj\rho(t')\bigr)\ket{b\bar a}\\
    =& \sum_{c\bar a} \sigma_{cb}(t')\frac{\rho_{\bar a\bar a}^\systb}{Z_\systb}\bra{a\bar a}\hat Ve^{-i\hat H_0(t-t')}\hat V\ket{c\bar a}e^{iE_b^\systa(t-t')}e^{iE_{\bar a}^\systb(t-t')}
\end{split}
\end{align}
Inserting a resolution of identity $1=\sum_{d\bar b}\ket{d\bar b}\bra{d\bar b}$, we get
\begin{align}
\begin{split}
    & \sum_{cd\bar a\bar b} \sigma_{cb}(t')\frac{\rho_{\bar a\bar a}^\systb}{Z_\systb}\bra{a\bar a}\hat V\ket{d\bar b}\bra{d\bar b}\hat V\ket{c\bar a}e^{i\omega_{bd}^\systa(t-t')}e^{i\omega_{\bar a\bar b}^\systb(t-t')}
\end{split}
\end{align}
where $\omega_{ab}^\systa$ are energy differences for the molecule, while $\omega_{\bar a\bar b}^\systb$ are energy differences for the environment. Similarly for the remaining terms in eq \ref{eq:double_commutator_written_out}, we get
\begin{align}
\begin{split}
    Z_\systb^{-1}e^{-i\omega_{ab}t}\bra{a\bar a}\pproj\comm{\tilde V(t)}{\comm{\tilde V(t')}{&\pproj\tilde\rho(t')}}\ket{b\bar a}dt' \\
    = \sum_{{cd\bar a\bar d}}\Bigl(
     &\sigma_{cb}(t')\bra{a\bar a}\hat V\ket{d\bar d}\bra{d\bar d}\hat V\ket{c\bar a}e^{i\omega_{bd}^\systa(t-t')}e^{i\omega_{\bar a\bar d}^\systb(t-t')}Z_\systb^{-1}\rho_{\bar a\bar a}^\systb\\
    -&\sigma_{cd}(t')\bra{d\bar d}\hat V\ket{b\bar a}\bra{a\bar a}\hat V\ket{c\bar d}e^{i\omega_{bc}^\systa(t-t')}e^{i\omega_{\bar a\bar d}^\systb(t-t')}Z_\systb^{-1}\rho_{\bar d\bar d}^\systb\\
    -&\sigma_{cd}(t')\bra{d\bar d}\hat V\ket{b\bar a}\bra{a\bar a}\hat V\ket{c\bar d}e^{i\omega_{da}^\systa(t-t')}e^{i\omega_{\bar d \bar a}^\systb(t-t')}Z_\systb^{-1}\rho_{\bar d\bar d}^\systb\\
    +&\sigma_{ad}(t)\bra{d\bar a}\hat V\ket{c\bar d}\bra{c\bar d}\hat V\ket{b\bar a}e^{i\omega_{ca}^\systa(t-t')}e^{i\omega_{\bar d\bar a}^\systb(t-t')}Z_\systb^{-1}\rho_{\bar a\bar a}^\systb\Bigr)
\end{split}
\end{align}
Now we must evaluate $\bra{a\bar a}\hat V\ket{b\bar b}\bra{c\bar b}\hat V\ket{d\bar a}$. By observing that the effective interaction operator
\begin{align}
    \hat V = \sum_{p\bar p}(V_{p\bar p}\cre{p}\ann{\bar p}+V_{p\bar p}^*\cre{\bar p}\ann{p})
\end{align}
changes the particle number by either $+1$ or $-1$ in each subsystem, we get two non-zero terms
\begin{align}
\begin{split}
\label{eq:aVbcVd}
    \bra{a\bar a}\hat V\ket{b\bar b}\bra{c\bar b}\hat V\ket{d\bar a} &= 
    \sum_{p\bar p q\bar q}V_{p\bar p}V_{q\bar q}^*\bra{a\bar a}\cre{p}\ann{\bar p}\ket{b\bar b}\bra{c\bar b}\cre{\bar q}\ann{q}\ket{d\bar a}\\
    &+ \sum_{p\bar p q\bar q}V_{p\bar p}^*V_{q\bar q}\bra{a\bar a}\cre{\bar p}\ann{p}\ket{b\bar b}\bra{c\bar b}\cre{q}\ann{\bar q}\ket{d\bar a}
\end{split}
\end{align}
Inserting expressions for the FCI states, we get for the first of these terms
\begin{align}
\begin{split}
    &\bra{a\bar a}\cre{p}\ann{\bar p}\ket{b\bar b}\bra{c\bar b}\cre{\bar q}\ann{q}\ket{d\bar a}\\
    =& \sum_{\substack{IJKL\\\bar I\bar J\bar K\bar L}}
    C_{Ia}^{\systa *}C_{\bar I\bar a}^{\systb *}C_{Jb}^\systa C_{\bar J\bar b}^\systb C_{Kc}^{\systa*} C_{\bar K\bar d}^{\systb*} C_{Ld}^\systa C_{\bar L\bar a}^\systb
    \vacbra \hat{A}_{\bar I}\hat{A}_I\cre{p}\ann{\bar p}\hat{A}_J^\dagger\hat{A}_{\bar J}^\dagger \vacket\vacbra\hat B_{\bar K}\hat{A}_{K}\cre{\bar q}\ann{q}\hat{A}_L^\dagger\hat{A}_{\bar L}^\dagger\vacket
\end{split}
\end{align}
We anticommute the creation and annihilation operators referring to the orbital space of the environment to the right and keep track of the number of anticommutations performed to get
\begin{align}
    \sum_{\substack{IJKL\\\bar I\bar J\bar K\bar L}}
    C_{Ia}^{\systa *}C_{\bar I\bar a}^{\systb *}C_{Jb}^\systa C_{\bar J\bar b}^\systb C_{Kc}^{\systa*} C_{\bar K\bar d}^{\systb*} C_{Ld}^\systa C_{\bar L\bar a}^\systb
    \vacbra 
    \hat{A}_I\cre{p}\hat{A}_J^\dagger\hat{A}_{\bar I}\ann{\bar p}\hat{A}_{\bar J}^\dagger 
    \vacket
    \vacbra \hat{A}_{K}\ann{q}\hat{A}_L^\dagger \hat B_{\bar K}\cre{\bar q}\hat{A}_{\bar L}^\dagger
    \vacket
    (-1)^{\Delta N_{bc}}
\end{align}
where $\Delta N_{bc}$ is the difference in the number of electrons between molecule states $\ket{b}$ and $\ket{c}$. Inserting the resolution of identity makes us write $\vacbra\hat A_I \cre{p} \hat A_J^\dagger \hat{A}_{\bar I}\ann{\bar p}\hat B_{\bar J}^\dagger\vacket = \vacbra\hat A_I \cre{p} \hat A_J^\dagger\vacket\vacbra \hat{A}_{\bar I}\ann{\bar p}\hat B_{\bar J}^\dagger\vacket$. This gives us
\begin{align}
\begin{split}
    \sum_{p\bar p q\bar q}V_{p\bar p}V_{q\bar q}^*\bra{a\bar a}\cre{p}\ann{\bar p}\ket{b\bar b}\bra{c\bar b}\cre{\bar q}\ann{q}\ket{d\bar a} = (-1)^{\Delta N_{bc}}
    \sum_{pq}\bra{a}\cre{p}\ket{b}\bra{c}\ann{q}\ket{d}\sum_{\bar p\bar q}V_{p\bar p}V_{q\bar q}^*\bra{\bar a}\ann{\bar p}\ket{\bar b}\bra{\bar b}\cre{\bar q}\ket{\bar a}
\end{split}
\end{align}
The same steps can be followed for the other term in eq \eqref{eq:aVbcVd}. The result becomes
\begin{align}
\begin{split}
    \bra{a\bar a}\hat V\ket{b\bar b}\bra{c\bar b}\hat V\ket{d\bar a} &= 
    (-1)^{\Delta N_{bc}}
    \sum_{pq}\bra{a}\cre{p}\ket{b}\bra{c}\ann{q}\ket{d}\sum_{\bar p\bar q}V_{p\bar p}V_{q\bar q}^*\bra{\bar a}\ann{\bar p}\ket{\bar b}\bra{\bar b}\cre{\bar q}\ket{\bar a}\\
    &+ (-1)^{\Delta N_{bc}}\sum_{pq}
        \bra{a}\ann{q}\ket{b}\bra{c}\cre{\bar p}\ket{d}\sum_{\bar p\bar q}V_{p\bar p}V_{q\bar q}^*\bra{\bar a}\cre{\bar q}\ket{\bar b}\bra{\bar b}\ann{\bar p}\ket{\bar a}
\end{split}
\end{align}
We have now separated the systems $\systa$ and $\systb$. Using these results and insertinh back into eq \eqref{eq:double_commutator_written_out}, we get the equation of motion for the reduced density matrix
\begin{align}
\label{eq:derivation_where_we_left_of}
\begin{split}
    \ddt\sigma_{ab}(t) = &-i\omega_{ab}\sigma_{ab}(t) - \sum_{cd}\sum_{pq}\int_{0}^t\Bigl(\\
    &\bigl(
        \bra{a}\cre{p}\ket{d}\bra{d}\ann{q}\ket{c}C_{pq}(t-t') + \bra{a}\ann{q}\ket{d}\bra{d}\cre{p}\ket{c}D_{pq}(t-t')
    \bigr)e^{i\omega_{bd} (t-t')}\sigma_{cb}(t')\\
    +&\bigl(
        \bra{c}\cre{q}\ket{d}\bra{d}\ann{p}\ket{b}C_{pq}(t-t')^* + \bra{c}\ann{p}\ket{d}\bra{d}\cre{q}\ket{b}D_{pq}(t-t')^*
    \bigr)e^{i\omega_{da} (t-t')}\sigma_{ac}(t')\\
    -&\bigl(
        \bra{d}\cre{p}\ket{b}\bra{a}\ann{q}\ket{c}C_{pq}(t-t') + \bra{d}\ann{q}\ket{b}\bra{a}\cre{p}\ket{c}D_{pq}(t-t')
    \bigr)e^{i\omega_{da} (t-t')}(-1)^{\Delta N_{cd}}\sigma_{cd}(t')\\
    -&\bigl(
        \bra{d}\cre{q}\ket{b}\bra{a}\ann{p}\ket{c}C_{pq}(t-t')^* + \bra{d}\ann{p}\ket{b}\bra{a}\cre{q}\ket{c}D_{pq}(t-t')^*
    \bigr)e^{i\omega_{bc} (t-t')}(-1)^{\Delta N_{cd}}\sigma_{cd}(t')
    \Bigr)dt'
\end{split}
\end{align}
where we have collected all the information about the environment in the correlation functions $C_{pq}(\tau)$ and $D_{pq}(\tau)$. All energy differences in this expression refer to the molecular system $\systa$, and the superscript will henceforth be dropped. The correlation functions are given by
\begin{align}
    C_{pq}(\tau) &= \sum_{\bar p\bar q}V_{p\bar p}V_{q\bar q}^*\sum_{\bar b}\bra{\bar b}\cre{\bar q}\Bigl(\sum_{\bar a}\frac{\rho_{\bar a\bar a}^\systb}{Z_\systb}\ket{\bar a}\bra{\bar a}\Bigr)\tann{\bar p}(\tau)\ket{\bar b}\\
    D_{pq}(\tau) &= \sum_{\bar p\bar q}V_{p\bar p}V_{q\bar q}^*\sum_{\bar b}\bra{\bar b}\ann{\bar p}\Bigl(\sum_{\bar a}\frac{\rho_{\bar a\bar a}^\systb}{Z_\systb}\ket{\bar a}\bra{\bar a}\Bigr)\tcre{\bar q}(\tau)\ket{\bar b}
\end{align}
We want to rewrite the above expression to the standard form of two-time cross-correlation functions. We do this by first redefining the environment density operator 
\begin{align}
    \rho_\systb=\sum_{\bar a}\frac{\rho_{\bar a\bar a}^\systb}{Z_\systb}\ket{\bar a}\bra{\bar a}
\end{align}
By using the definition of the partial trace, we may write
\begin{align}
    C_{pq}(\tau) &= \sum_{\bar p\bar q}V_{p\bar p}V_{q\bar q}^*\Tr{\systb}{\cre{\bar q}\rho_\systb\tann{\bar p}(\tau)}\\
    D_{pq}(\tau) &= \sum_{\bar p\bar q}V_{p\bar p}V_{q\bar q}^*\Tr{\systb}{\ann{\bar p}\rho_\systb\tcre{\bar q}(\tau)}
\end{align}
By using the cyclic property of the trace, and writing $\langle\Omega\rangle_\systb\equiv\Tr{\systb}{\Omega\rho_\systb}$ we end up with
\begin{align}
    C_{pq}(\tau) &= \sum_{\bar p\bar q}V_{p\bar p}V_{q\bar q}^*\bigl\langle\tann{\bar p}(\tau)\cre{\bar q}\bigr\rangle_\systb\\
    D_{pq}(\tau) &= \sum_{\bar p\bar q}V_{p\bar p}V_{q\bar q}^*\bigl\langle\tcre{\bar q}(\tau)\ann{\bar p}\bigr\rangle_\systb
\end{align}
We note the symmetry of the correlation functions $C_{pq}(\tau)^*=C_{qp}(-\tau)$ and $D_{pq}(\tau)^*=D_{qp}(-\tau)$. This makes us write eq \eqref{eq:derivation_where_we_left_of}, with $\tau=t-t'$, as
\begin{align}
\begin{split}
    \ddt\sigma_{ab}(t) = &-i\omega_{ab}\sigma_{ab}(t) - \sum_{cd}\sum_{pq}\int_{0}^t\Bigl(\\
    &\bigl(
        \bra{a}\cre{p}\ket{d}\bra{d}\ann{q}\ket{c}C_{pq}(\tau) + \bra{a}\ann{q}\ket{d}\bra{d}\cre{p}\ket{c}D_{pq}(\tau)
    \bigr)e^{i\omega_{bd}\tau}\sigma_{cb}(t-\tau)\\
    +&\bigl(
        \bra{c}\cre{p}\ket{d}\bra{d}\ann{q}\ket{b}C_{pq}(-\tau) + \bra{c}\ann{q}\ket{d}\bra{d}\cre{p}\ket{b}D_{pq}(-\tau)
    \bigr)e^{i\omega_{da}\tau}\sigma_{ac}(t-\tau)\\
    -&\bigl(
        \bra{d}\cre{p}\ket{b}\bra{a}\ann{q}\ket{c}C_{pq}(\tau) + \bra{d}\ann{q}\ket{b}\bra{a}\cre{p}\ket{c}D_{pq}(\tau)
    \bigr)e^{i\omega_{da}\tau}(-1)^{\Delta N_{cd}}\sigma_{cd}(t-\tau)\\
    -&\bigl(
        \bra{d}\cre{p}\ket{b}\bra{a}\ann{q}\ket{c}C_{pq}(-\tau) + \bra{d}\ann{q}\ket{b}\bra{a}\cre{p}\ket{c}D_{pq}(-\tau)
    \bigr)e^{i\omega_{bc}\tau}(-1)^{\Delta N_{cd}}\sigma_{cd}(t-\tau)
    \Bigr)d\tau
\end{split}
\end{align}
We now apply two approximations:
\begin{enumerate}
    \item The Markov approximation
    \item The correlation functions go to $0$ for large $\tau$
\end{enumerate}
The first approximation makes $\sigma(t-\tau)\to\sigma(t)$, resulting in a differential equation local in time. This is the Markov approximation saying that the system's dynamics only depend on the current state, and carry no information about earlier times. The second approximation says that $C_{pq}(\tau)\to0$ and $D_{pq}(\tau)\to0$ for large time differences $\tau$. This is justified by assuming that the relaxation time of the environment is small. If the correlation functions drop to zero quickly enough, the upper integration limit can be approximately extended to $\infty$. By introducing the tensors
\begin{align}
\begin{split}
    \Gamma_{abcd}^x(\omega) 
    &= 
    \sum_{pq}\bra{a}\cre{p}\ket{b}\bra{c}\ann{q}\ket{d}\int_0^\infty e^{i\omega\tau}C_{pq}(x\tau)d\tau\\
    &+
    \sum_{pq}\bra{a}\ann{q}\ket{b}\bra{c}\cre{p}\ket{d}\int_0^\infty e^{i\omega\tau}D_{pq}(x\tau)d\tau
\end{split}
\end{align}
with $x=+,-$, we can write the equation as
\begin{align}
\begin{split}
    \ddt\sigma_{ab}(t) = -i\omega_{ab}\sigma_{ab}(t) + \sum_{cd}\Bigl(
        &\Gamma_{dbac}^+(\omega_{da}) + \Gamma_{dbac}^-(\omega_{bc})\\
        - 
        \delta_{bd}\sum_e&\Gamma_{aeec}^+(\omega_{de})-\delta_{ac}\sum_e\Gamma_{deeb}^-(\omega_{ec})
    \Bigr)(-1)^{\Delta N_{cd}}\sigma_{cd}(t)
\end{split}
\end{align}
The $+/-$ sign of $\boldsymbol\Gamma^x$ specifies the sign of the time difference in the correlation functions and reflects the different Liouville space pathways of forward and backward time propagation. By defining the Redfield tensor $\mathbf{R}$ as
\begin{align}
    R_{abcd} = \Bigl(\Gamma_{dbac}^+(\omega_{da}) + \Gamma_{dbac}^-(\omega_{bc})
        - 
        \delta_{bd}\sum_e\Gamma_{aeec}^+(\omega_{de})-\delta_{ac}\sum_e\Gamma_{deeb}^-(\omega_{ec})
    \Bigr)(-1)^{\Delta N_{cd}}
\end{align}
the Redfield equation can finally be written as
\begin{align}
    \ddt\sigma_{ab}(t) = -i\omega_{ab}\sigma_{ab}(t) + \sum_{cd}R_{abcd}\sigma_{cd}(t)
\end{align}

\section{Simplifying the correlation function}
\label{sec:correlation_function}

We have the bath time-correlation functions

\begin{align}
    C_{pq}(\tau) &= \sum_{\bar p\bar q}\sum_{\bar p\bar q}V_{p\bar p}V_{q\bar q}^*\frac{\rho_{\bar p\bar p}^\systb}{\rho^\systb}e^{i\omega_{\bar p\bar q}^\systb\tau}\bra{\bar p}\ann{\bar p}\ket{\bar q}\bra{\bar q}\cre{\bar q}\ket{\bar p}\\
    D_{pq}(\tau) &= \sum_{\bar p\bar q}\sum_{\bar p\bar q}V_{p\bar p}V_{q\bar q}^*\frac{\rho_{\bar p\bar p}^\systb}{\rho^\systb}e^{i\omega_{\bar p\bar q}^\systb\tau}\bra{\bar p}\cre{\bar q}\ket{\bar q}\bra{\bar q}\ann{\bar p}\ket{\bar p}
\end{align}
The environment is assumed to be large. We approximate it to have a continuous spectrum, and that its states are single determinants. We then rewrite the correlation functions as
\begin{align}
    C_{pq}(\tau) &= \sum_{\bar p\bar q}\int d\omega_1 \int d\omega_2V_{p}(\omega_1)V_{q}(\omega_2)^*\frac{\rho_{\bar p\bar p}^\systb}{\rho^\systb}e^{i\omega_{\bar p\bar q}^\systb\tau}\bra{\bar p}\ann{\omega_1}\ket{\bar q}\bra{\bar q}\cre{\omega_2}\ket{\bar p}\\
    D_{pq}(\tau) &= \sum_{\bar p\bar q}\int d\omega_1 \int d\omega_2V_{p}(\omega_1)V_{q}(\omega_2)^*\frac{\rho_{\bar p\bar p}^\systb}{\rho^\systb}e^{i\omega_{\bar p\bar q}^\systb\tau}\bra{\bar p}\cre{\omega_2}\ket{\bar q}\bra{\bar q}\ann{\omega_1}\ket{\bar p}
\end{align}
We simplify
\begin{align}
\begin{split}
    \sum_{\bar q}e^{i\omega_{\bar p\bar q}^\systb\tau}\bra{\bar p}\ann{\omega_1}\ket{\bar q}\bra{\bar q}\cre{\omega_2}\ket{\bar p} 
    &= 
    \sum_{\bar q}\bra{\bar p}e^{iH_\systb\tau}\ann{\omega_1}e^{-iH_\systb\tau}\ket{\bar q}\bra{\bar q}\cre{\omega_2}\ket{\bar p}\\
    &= 
    \bra{\bar p}e^{iH_\systb\tau}\ann{\omega_1}e^{-iH_\systb\tau}\cre{\omega_2}\ket{\bar p}\\
    &= 
    \bra{\bar p}e^{iE_{\bar p}^\systb\tau}\ann{\omega_1}e^{-i(E_{\bar p}^\systb + \omega_2)\tau}\ket{\bar p+\omega_{2}}[1-k_{\bar p}(\omega_2)]\\
    &= 
    \bra{\bar p}\ann{\omega_1}\ket{\bar p+\omega_{2}}e^{-i\omega_2\tau}[1-k_{\bar p}(\omega_2)]\\
    &= 
    \delta(\omega_1-\omega_2)e^{-i\omega_2\tau}[1-k_{\bar p}(\omega_2)]
\end{split}
\end{align}
where $k_{\bar p}(\omega)$ is the occupation number at energy $\omega$ for state $\ket{\psi_{\bar p}^\systb}$. Similarly, we have
\begin{align}
\begin{split}
    \sum_{\bar q}e^{i\omega_{\bar p\bar q}^\systb\tau}\bra{\bar p}\cre{\omega_2}\ket{\bar q}\bra{\bar q}\ann{\omega_1}\ket{\bar p}
    &=
    \bra{\bar p}e^{iH_\systb\tau}\cre{\omega_2}e^{-iH_\systb\tau}\ann{\omega_1}\ket{\bar p}\\
    &=
    \bra{\bar p}\cre{\omega_2}\ann{\omega_1}\ket{\bar p-\omega_1}e^{i\omega_1\tau}k_{\bar p}(\omega_1)\\
    &=
    \delta(\omega_1-\omega_2)e^{i\omega_1\tau}k_{\bar p}(\omega_1)
\end{split}
\end{align}
Since $\rho^\systb$ is defined as $\sum_{\bar p}\rho_{\bar p\bar p}^\systb$ we can write
\begin{align}
    \sum_{\bar p}\frac{\rho_{\bar p\bar p}^\systb}{\rho^\systb}k_{\bar p}(\omega) = \sum_{\bar p}\frac{\rho_{\bar p\bar p}k_{\bar p}(\omega)}{\sum_{\bar q}\rho_{\bar q\bar q}^\systb} \equiv \bar{n}(\omega)
\end{align}
\begin{align}
    \sum_{\bar p}\frac{\rho_{\bar p\bar p}^\systb}{\rho^\systb}[1-k_{\bar p}(\omega)] = \frac{\sum_{\bar p}\rho_{\bar p\bar p}^\systb}{\sum_{\bar q}\rho_{\bar q\bar q}^\systb} - \sum_{\bar p}\frac{\rho_{\bar p\bar p}k_{\bar p}(\omega)}{\sum_{\bar q}\rho_{\bar q\bar q}^\systb} = 1 - \bar{n}(\omega)
\end{align}
where $\bar n(\omega)$ is the expected number of electrons in the environment at energy $\omega$, a number between $0$ and $1$. Correspondingly, $[1-\bar n(\omega)]$ is the expected number of holes at energy $\omega$. The function $\bar n(\omega)$ could for instance be described by the Fermi-Dirac function given a temperature $T$ and a chemical potential $p$,
\begin{align}
    \bar n(\omega; p, T) = \Bigl[\exp\Bigl(\frac{\omega-p}{k_BT}\Bigr)+1\Bigr]^{-1}
\end{align}
where $k_B$ is the Boltzmann constant.
The time-correlation functions are then
\begin{align}
    \label{eq:correlation_function_minus_simplified}
    C_{pq}(\tau) &= \int d\omega V_{p}(\omega)V_{q}(\omega)^*e^{-i\omega\tau}[1-\bar n(\omega)]\\
    \label{eq:correlation_function_plus_simplified}
    D_{pq}(\tau) &= \int d\omega V_p(\omega)V_q(\omega)^*e^{i\omega\tau}\bar n(\omega)
\end{align}

\subsection{The secular approximation}

Assuming fast-oscillating and decaying off-diagonal elements of the reduced density operator, the Redfield equation for the populations reduces to
\begin{align}
    \ddt\sigma_{aa}(t) &= \sum_b k_{a\leftarrow b}\sigma_{bb}(t)
    - \sum_b k_{b\leftarrow a}\sigma_{aa}(t)
\end{align}
with the population transfer rates $k_{b\leftarrow a} = \Gamma_{abba}^+(\omega_{ab}) + \Gamma_{abba}^-(\omega_{ba})$. We evaluate the rates to be
\begin{align}
\begin{split}
    k_{b\leftarrow a} = \sum_{pq}\bigl(
        &
        \bra{a}\cre{p}\ket{b}\bra{b}\ann{q}\ket{a}\int_0^\infty  e^{i\omega_{ab}\tau}C_{pq}(\tau)d\tau\\
        +&
        \bra{a}\ann{q}\ket{b}\bra{b}\cre{p}\ket{a}\int_0^\infty  e^{i\omega_{ab}\tau}D_{pq}(\tau)d\tau\\
        +&
        \bra{a}\cre{p}\ket{b}\bra{b}\ann{q}\ket{a}\int_0^\infty  e^{i\omega_{ba}\tau}C_{pq}(-\tau)d\tau\\
        +&
        \bra{a}\ann{q}\ket{b}\bra{b}\cre{p}\ket{a}\int_0^\infty  e^{i\omega_{ba}\tau}D_{pq}(-\tau)d\tau
    \bigr)\\
    = \sum_{pq}\bigl(
        &
        \bra{a}\cre{p}\ket{b}\bra{b}\ann{q}\ket{a}\int_{-\infty}^\infty  e^{i\omega_{ab}\tau}C_{pq}(\tau)d\tau\\
        +&
        \bra{a}\ann{q}\ket{b}\bra{b}\cre{p}\ket{a}\int_{-\infty}^\infty  e^{i\omega_{ab}\tau}D_{pq}(\tau)d\tau
    \bigr)
\end{split}
\end{align}
Inserting the simplified time-correlation functions of eqs \eqref{eq:correlation_function_minus_simplified} - \eqref{eq:correlation_function_plus_simplified}, we get
\begin{align}
\label{eq:SI_transition_rate_final_expression}
\begin{split}
    k_{b\leftarrow a} = \sum_{pq}\Bigl(
        &
        \bra{a}\cre{p}\ket{b}\bra{b}\ann{q}\ket{a} \int d\omega V_{p}(\omega)V_{q}(\omega)^*[1-\bar n(\omega)]\int_{-\infty}^\infty d\tau e^{i(\omega_{ab}-\omega)\tau}\\
        +&
        \bra{a}\ann{q}\ket{b}\bra{b}\cre{p}\ket{a}\int d\omega V_p(\omega)V_q(\omega)^*\bar n(\omega)\int_{-\infty}^\infty d\tau e^{i(\omega_{ab}+\omega)\tau}
    \Bigr)\\
    = 2\pi\sum_{pq}\Bigl(
        &
        \bra{a}\cre{p}\ket{b}\bra{b}\ann{q}\ket{a} \int d\omega V_{p}(\omega)V_{q}(\omega)^*[1-\bar n(\omega)]\delta(\omega_{ab}-\omega)\\
        +&
        \bra{a}\ann{q}\ket{b}\bra{b}\cre{p}\ket{a}\int d\omega V_p(\omega)V_q(\omega)^*\bar n(\omega) \delta(\omega_{ab}+\omega)
    \Bigr)\\
    = 2\pi\sum_{pq}\Bigl(
        &
        \bra{a}\cre{p}\ket{b}\bra{b}\ann{q}\ket{a} V_{p}(\omega_{ab})V_{q}(\omega_{ab})^*[1-\bar n(\omega_{ab})]\\
        +&
        \bra{a}\ann{q}\ket{b}\bra{b}\cre{p}\ket{a} V_p(\omega_{ba})V_q(\omega_{ba})^*\bar n(\omega_{ba})
    \Bigr)\\
    = \sum_{pq}\Bigl(
        &
        \bra{a}\cre{p}\ket{b}\bra{b}\ann{q}\ket{a} J_{pq}(\omega_{ab})[1-\bar n(\omega_{ab})]\\
        +&
        \bra{a}\ann{q}\ket{b}\bra{b}\cre{p}\ket{a} J_{pq}(\omega_{ba})\bar n(\omega_{ba})
    \Bigr)
\end{split}
\end{align}
where we have introduced the spectral density function $J_{pq}(\omega)=2\pi V_p(\omega)V_q(\omega)^*$ which carries information about the density of states of the environment and the interaction strengths.

\subsection{Energy level broadening}
\label{sec:SI_energy_level_broadening}

With broadened energy levels due to finite lifetimes of excited states, the Dirac delta functions containing excitation energies in eq \eqref{eq:SI_transition_rate_final_expression} are replaced by normalized Gaussian functions with standard deviation $\gamma$

\begin{align}
\label{eq:energy_level_broadening}
    \delta(\omega_{ab} \pm \omega) \to \frac{1}{\sqrt{2\pi\gamma^2}}\exp\Bigl(-\frac{(\omega_{ab}\pm\omega)^2}{2\gamma^2}\Bigr)
\end{align}
such that the population transfer rates for single-determinantal system states become
\begin{align}
\label{eq:transition_rate_with_energy_level_broadening}
\begin{split}
    k_{b\leftarrow a}
    = &\frac{1}{\sqrt{2\pi\gamma^2}}\sum_{p}\Bigl(\\
    &\bra{a}\cre{p}\ket{b}\bra{b}\ann{p}\ket{a}\int_{-\infty}^\infty d\omega J_{p}(\omega)\exp\Bigl(-\frac{(\omega_{ab}\pm\omega)^2}{2\gamma^2}\Bigr)[1-\bar n(\omega)]\\
        +&
        \bra{a}\ann{p}\ket{b}\bra{b}\cre{p}\ket{a}\int_{-\infty}^\infty d\omega J_{p}(\omega)\exp\Bigl(-\frac{(\omega_{ba}\pm\omega)^2}{2\gamma^2}\Bigr)\bar n(\omega)
    \Bigr)
\end{split}
\end{align}

\section*{Benzene geometry}

The geometry of the benzene molecule is shown in Table \ref{tab:benzene_geometry}.

\begin{table}[H]
    \centering
    \begin{tabular}{c c c c}
    \toprule
    Atom & $x[\si{\angstrom}]$ & $y[\si{\angstrom}]$ & $z[\si{\angstrom}]$\\
    \midrule
    C1  &  -0.695000000  &    1.203775311     &    0.000000000\\
    C2  &  0.695000000   &    1.203775311     &    0.000000000\\
    C3  &  1.390000000   &    0.000000000     &    0.000000000\\
    C4  &  0.695000000   &    -1.203775311    &    0.000000000\\
    C5  &  -0.695000000  &    -1.203775311    &    0.000000000\\
    C6  &  -1.390000000  &    0.000000000     &    0.000000000\\
    H1  &  1.240000000   &    2.147743001     &    0.000000000\\
    H2  &  -1.240000000  &    -2.147743001    &    0.000000000\\
    H3  &  -1.240000000  &    2.147743001     &    0.000000000\\
    H4  &  1.240000000   &    -2.147743001    &    0.000000000\\
    H5  &  2.480000000   &    0.000000000     &    0.000000000\\
    H6  &  -2.480000000  &    0.000000000     &    0.000000000\\
    \bottomrule
    \end{tabular}
    \caption{The benzene geometry with C-C bond lengths of \SI{1.39}{\angstrom}, C-H bond lengths of \SI{1.09}{\angstrom}, and all angles equal to $120^\circ$.}
    \label{tab:benzene_geometry}
\end{table}

%% file: main.bbl
\providecommand{\noopsort}[1]{}\providecommand{\singleletter}[1]{#1}%
\providecommand{\latin}[1]{#1}
\makeatletter
\providecommand{\doi}
  {\begingroup\let\do\@makeother\dospecials
  \catcode`\{=1 \catcode`\}=2 \doi@aux}
\providecommand{\doi@aux}[1]{\endgroup\texttt{#1}}
\makeatother
\providecommand*\mcitethebibliography{\thebibliography}
\csname @ifundefined\endcsname{endmcitethebibliography}
  {\let\endmcitethebibliography\endthebibliography}{}
\begin{mcitethebibliography}{86}
\providecommand*\natexlab[1]{#1}
\providecommand*\mciteSetBstSublistMode[1]{}
\providecommand*\mciteSetBstMaxWidthForm[2]{}
\providecommand*\mciteBstWouldAddEndPuncttrue
  {\def\EndOfBibitem{\unskip.}}
\providecommand*\mciteBstWouldAddEndPunctfalse
  {\let\EndOfBibitem\relax}
\providecommand*\mciteSetBstMidEndSepPunct[3]{}
\providecommand*\mciteSetBstSublistLabelBeginEnd[3]{}
\providecommand*\EndOfBibitem{}
\mciteSetBstSublistMode{f}
\mciteSetBstMaxWidthForm{subitem}{(\alph{mcitesubitemcount})}
\mciteSetBstSublistLabelBeginEnd
  {\mcitemaxwidthsubitemform\space}
  {\relax}
  {\relax}

\bibitem[Höfener and Visscher(2012)Höfener, and Visscher]{hofener2012wfinwf}
Höfener,~S.; Visscher,~L. Calculation of electronic excitations using
  wave-function in wave-function frozen-density embedding. \emph{The Journal of
  Chemical Physics} \textbf{2012}, \emph{137}, 204120\relax
\mciteBstWouldAddEndPuncttrue
\mciteSetBstMidEndSepPunct{\mcitedefaultmidpunct}
{\mcitedefaultendpunct}{\mcitedefaultseppunct}\relax
\EndOfBibitem
\bibitem[S{\ae}ther \latin{et~al.}(2017)S{\ae}ther, Kj{\ae}rgaard, Koch, and
  H{\o}yvik]{saether2017mlhf}
S{\ae}ther,~S.; Kj{\ae}rgaard,~T.; Koch,~H.; H{\o}yvik,~I.-M. Density-based
  multilevel Hartree--Fock model. \emph{Journal of Chemical Theory and
  Computation} \textbf{2017}, \emph{13}, 5282--5290\relax
\mciteBstWouldAddEndPuncttrue
\mciteSetBstMidEndSepPunct{\mcitedefaultmidpunct}
{\mcitedefaultendpunct}{\mcitedefaultseppunct}\relax
\EndOfBibitem
\bibitem[Myhre \latin{et~al.}(2014)Myhre, S{\'a}nchez~de Mer{\'a}s, and
  Koch]{myhre2014mlcc}
Myhre,~R.~H.; S{\'a}nchez~de Mer{\'a}s,~A.~M.; Koch,~H. Multi-level coupled
  cluster theory. \emph{The Journal of chemical physics} \textbf{2014},
  \emph{141}\relax
\mciteBstWouldAddEndPuncttrue
\mciteSetBstMidEndSepPunct{\mcitedefaultmidpunct}
{\mcitedefaultendpunct}{\mcitedefaultseppunct}\relax
\EndOfBibitem
\bibitem[Jacob and Neugebauer(2014)Jacob, and Neugebauer]{jacob2014subsystem}
Jacob,~C.~R.; Neugebauer,~J. Subsystem density-functional theory. \emph{WIREs
  Computational Molecular Science} \textbf{2014}, \emph{4}, 325–362\relax
\mciteBstWouldAddEndPuncttrue
\mciteSetBstMidEndSepPunct{\mcitedefaultmidpunct}
{\mcitedefaultendpunct}{\mcitedefaultseppunct}\relax
\EndOfBibitem
\bibitem[Lee \latin{et~al.}(2019)Lee, Welborn, Manby, and
  Miller]{lee2019projection}
Lee,~S. J.~R.; Welborn,~M.; Manby,~F.~R.; Miller,~T. F.~I. Projection-Based
  Wavefunction-in-DFT Embedding. \emph{Accounts of Chemical Research}
  \textbf{2019}, \emph{52}, 1359–1368\relax
\mciteBstWouldAddEndPuncttrue
\mciteSetBstMidEndSepPunct{\mcitedefaultmidpunct}
{\mcitedefaultendpunct}{\mcitedefaultseppunct}\relax
\EndOfBibitem
\bibitem[Manby \latin{et~al.}(2012)Manby, Stella, Goodpaster, and
  Miller]{manby2012simple}
Manby,~F.~R.; Stella,~M.; Goodpaster,~J.~D.; Miller,~T. F.~I. A Simple, Exact
  Density-Functional-Theory Embedding Scheme. \emph{Journal of Chemical Theory
  and Computation} \textbf{2012}, \emph{8}, 2564–2568\relax
\mciteBstWouldAddEndPuncttrue
\mciteSetBstMidEndSepPunct{\mcitedefaultmidpunct}
{\mcitedefaultendpunct}{\mcitedefaultseppunct}\relax
\EndOfBibitem
\bibitem[Senn and Thiel(2009)Senn, and Thiel]{senn2009qm}
Senn,~H.~M.; Thiel,~W. QM/MM Methods for Biomolecular Systems. \emph{Angewandte
  Chemie International Edition} \textbf{2009}, \emph{48}, 1198–1229\relax
\mciteBstWouldAddEndPuncttrue
\mciteSetBstMidEndSepPunct{\mcitedefaultmidpunct}
{\mcitedefaultendpunct}{\mcitedefaultseppunct}\relax
\EndOfBibitem
\bibitem[Cappelli(2016)]{cappelli2016integrated}
Cappelli,~C. Integrated QM/polarizable MM/continuum approaches to model
  chiroptical properties of strongly interacting solute–solvent systems.
  \emph{International Journal of Quantum Chemistry} \textbf{2016}, \emph{116},
  1532–1542\relax
\mciteBstWouldAddEndPuncttrue
\mciteSetBstMidEndSepPunct{\mcitedefaultmidpunct}
{\mcitedefaultendpunct}{\mcitedefaultseppunct}\relax
\EndOfBibitem
\bibitem[Tzeliou \latin{et~al.}(2022)Tzeliou, Mermigki, and
  Tzeli]{tzeliou2022qmmmreview}
Tzeliou,~C.~E.; Mermigki,~M.~A.; Tzeli,~D. Review on the QM/MM Methodologies
  and Their Application to Metalloproteins. \emph{Molecules} \textbf{2022},
  \emph{27}, 2660\relax
\mciteBstWouldAddEndPuncttrue
\mciteSetBstMidEndSepPunct{\mcitedefaultmidpunct}
{\mcitedefaultendpunct}{\mcitedefaultseppunct}\relax
\EndOfBibitem
\bibitem[Marcus(1964)]{marcus1964chemical}
Marcus,~R.~A. Chemical and electrochemical electron-transfer theory.
  \emph{Annual review of physical chemistry} \textbf{1964}, \emph{15},
  155--196\relax
\mciteBstWouldAddEndPuncttrue
\mciteSetBstMidEndSepPunct{\mcitedefaultmidpunct}
{\mcitedefaultendpunct}{\mcitedefaultseppunct}\relax
\EndOfBibitem
\bibitem[Marcus and Sutin(1985)Marcus, and Sutin]{marcus1985electron}
Marcus,~R.~A.; Sutin,~N. Electron transfers in chemistry and biology.
  \emph{Biochimica et Biophysica Acta (BBA) - Reviews on Bioenergetics}
  \textbf{1985}, \emph{811}, 265–322\relax
\mciteBstWouldAddEndPuncttrue
\mciteSetBstMidEndSepPunct{\mcitedefaultmidpunct}
{\mcitedefaultendpunct}{\mcitedefaultseppunct}\relax
\EndOfBibitem
\bibitem[Newton(1991)]{newton1991quantum}
Newton,~M.~D. Quantum chemical probes of electron-transfer kinetics: the nature
  of donor-acceptor interactions. \emph{Chemical Reviews} \textbf{1991},
  \emph{91}, 767–792\relax
\mciteBstWouldAddEndPuncttrue
\mciteSetBstMidEndSepPunct{\mcitedefaultmidpunct}
{\mcitedefaultendpunct}{\mcitedefaultseppunct}\relax
\EndOfBibitem
\bibitem[Fletcher(2010)]{fletcher2010theory}
Fletcher,~S. The theory of electron transfer. \emph{Journal of Solid State
  Electrochemistry} \textbf{2010}, \emph{14}, 705–739\relax
\mciteBstWouldAddEndPuncttrue
\mciteSetBstMidEndSepPunct{\mcitedefaultmidpunct}
{\mcitedefaultendpunct}{\mcitedefaultseppunct}\relax
\EndOfBibitem
\bibitem[Blumberger(2015)]{blumberger2015recent}
Blumberger,~J. Recent Advances in the Theory and Molecular Simulation of
  Biological Electron Transfer Reactions. \emph{Chemical Reviews}
  \textbf{2015}, \emph{115}, 11191–11238\relax
\mciteBstWouldAddEndPuncttrue
\mciteSetBstMidEndSepPunct{\mcitedefaultmidpunct}
{\mcitedefaultendpunct}{\mcitedefaultseppunct}\relax
\EndOfBibitem
\bibitem[Dzhioev and Kosov(2012)Dzhioev, and Kosov]{dzhioev2012nonequilibrium}
Dzhioev,~A.~A.; Kosov,~D.~S. Nonequilibrium perturbation theory in
  Liouville–Fock space for inelastic electron transport. \emph{Journal of
  Physics: Condensed Matter} \textbf{2012}, \emph{24}, 225304\relax
\mciteBstWouldAddEndPuncttrue
\mciteSetBstMidEndSepPunct{\mcitedefaultmidpunct}
{\mcitedefaultendpunct}{\mcitedefaultseppunct}\relax
\EndOfBibitem
\bibitem[Galperin and Nitzan(2012)Galperin, and Nitzan]{Galperin:2012aa}
Galperin,~M.; Nitzan,~A. Molecular optoelectronics: the interaction of
  molecular conduction junctions with light. \emph{Physical Chemistry Chemical
  Physics} \textbf{2012}, \emph{14}, 9421--9438\relax
\mciteBstWouldAddEndPuncttrue
\mciteSetBstMidEndSepPunct{\mcitedefaultmidpunct}
{\mcitedefaultendpunct}{\mcitedefaultseppunct}\relax
\EndOfBibitem
\bibitem[Elenewski \latin{et~al.}(2017)Elenewski, Gruss, and
  Zwolak]{elenewski2017communication}
Elenewski,~J.~E.; Gruss,~D.; Zwolak,~M. Communication: Master equations for
  electron transport: The limits of the Markovian limit. \emph{The Journal of
  Chemical Physics} \textbf{2017}, \emph{147}, 151101\relax
\mciteBstWouldAddEndPuncttrue
\mciteSetBstMidEndSepPunct{\mcitedefaultmidpunct}
{\mcitedefaultendpunct}{\mcitedefaultseppunct}\relax
\EndOfBibitem
\bibitem[Philbin \latin{et~al.}(2021)Philbin, Levy, Narang, and
  Dou]{Philbin_Levy_Narang_Dou_2021}
Philbin,~J.~P.; Levy,~A.; Narang,~P.; Dou,~W. Asymmetric Spin Transport in
  Colloidal Quantum Dot Junctions. \emph{The Journal of Physical Chemistry C}
  \textbf{2021}, \emph{125}, 26661–26669\relax
\mciteBstWouldAddEndPuncttrue
\mciteSetBstMidEndSepPunct{\mcitedefaultmidpunct}
{\mcitedefaultendpunct}{\mcitedefaultseppunct}\relax
\EndOfBibitem
\bibitem[Naaman \latin{et~al.}(2022)Naaman, Waldeck, and
  Fransson]{Naaman:2022aa}
Naaman,~R.; Waldeck,~D.~H.; Fransson,~J. New Perspective on Electron Transfer
  through Molecules. \emph{The Journal of Physical Chemistry Letters}
  \textbf{2022}, \emph{13}, 11753--11759\relax
\mciteBstWouldAddEndPuncttrue
\mciteSetBstMidEndSepPunct{\mcitedefaultmidpunct}
{\mcitedefaultendpunct}{\mcitedefaultseppunct}\relax
\EndOfBibitem
\bibitem[Breuer and Petruccione(2002)Breuer, and
  Petruccione]{breuer_2002_theory_of_open_quantum_systems}
Breuer,~H.~P.; Petruccione,~F. \emph{The theory of open quantum systems};
  Oxford University Press: Great Clarendon Street, 2002\relax
\mciteBstWouldAddEndPuncttrue
\mciteSetBstMidEndSepPunct{\mcitedefaultmidpunct}
{\mcitedefaultendpunct}{\mcitedefaultseppunct}\relax
\EndOfBibitem
\bibitem[May and K{\"u}hn(2011)May, and
  K{\"u}hn]{may_2011_charge_and_energy_transfer_dynamics_in_molecular_systems}
May,~V.; K{\"u}hn,~O. \emph{Charge and Energy Transfer Dynamics in Molecular
  Systems}; John Wiley \& Sons: Hoboken, NJ, 2011\relax
\mciteBstWouldAddEndPuncttrue
\mciteSetBstMidEndSepPunct{\mcitedefaultmidpunct}
{\mcitedefaultendpunct}{\mcitedefaultseppunct}\relax
\EndOfBibitem
\bibitem[Nitzan(2013)]{nitzan_2013_chemical_dynamics_in_condensed_phases}
Nitzan,~A. \emph{Chemical Dynamics in Condensed Phases: Relaxation, Transfer,
  and Reactions in Condensed Molecular Systems}; OUP Oxford, 2013\relax
\mciteBstWouldAddEndPuncttrue
\mciteSetBstMidEndSepPunct{\mcitedefaultmidpunct}
{\mcitedefaultendpunct}{\mcitedefaultseppunct}\relax
\EndOfBibitem
\bibitem[Rivas and Huelga(2011)Rivas, and
  Huelga]{rivas_2011_open_quantum_systems_an_introduction}
Rivas,~{\'A}.; Huelga,~S. \emph{Open Quantum Systems: An Introduction};
  SpringerBriefs in Physics; Springer Berlin Heidelberg, 2011\relax
\mciteBstWouldAddEndPuncttrue
\mciteSetBstMidEndSepPunct{\mcitedefaultmidpunct}
{\mcitedefaultendpunct}{\mcitedefaultseppunct}\relax
\EndOfBibitem
\bibitem[Stefanucci and van Leeuwen(2013)Stefanucci, and van
  Leeuwen]{stefanucci_2013_nonequilibrium_many_body_theory}
Stefanucci,~G.; van Leeuwen,~R. \emph{Nonequilibrium Many-Body Theory of
  Quantum Systems: A Modern Introduction}; Cambridge University Press,
  2013\relax
\mciteBstWouldAddEndPuncttrue
\mciteSetBstMidEndSepPunct{\mcitedefaultmidpunct}
{\mcitedefaultendpunct}{\mcitedefaultseppunct}\relax
\EndOfBibitem
\bibitem[Yan and Mukamel(1988)Yan, and Mukamel]{Yan:1988aa}
Yan,~Y.~J.; Mukamel,~S. Electronic dephasing, vibrational relaxation, and
  solvent friction in molecular nonlinear optical line shapes. \emph{The
  Journal of Chemical Physics} \textbf{1988}, \emph{89}, 5160--5176\relax
\mciteBstWouldAddEndPuncttrue
\mciteSetBstMidEndSepPunct{\mcitedefaultmidpunct}
{\mcitedefaultendpunct}{\mcitedefaultseppunct}\relax
\EndOfBibitem
\bibitem[Khalil \latin{et~al.}(2004)Khalil, Demird{\"o}ven, and
  Tokmakoff]{Khalil:2004aa}
Khalil,~M.; Demird{\"o}ven,~N.; Tokmakoff,~A. Vibrational coherence transfer
  characterized with Fourier-transform 2D IR spectroscopy. \emph{The Journal of
  Chemical Physics} \textbf{2004}, \emph{121}, 362--373\relax
\mciteBstWouldAddEndPuncttrue
\mciteSetBstMidEndSepPunct{\mcitedefaultmidpunct}
{\mcitedefaultendpunct}{\mcitedefaultseppunct}\relax
\EndOfBibitem
\bibitem[Galperin \latin{et~al.}(2008)Galperin, Ratner, Nitzan, and
  Troisi]{Galperin:2008aa}
Galperin,~M.; Ratner,~M.~A.; Nitzan,~A.; Troisi,~A. Nuclear Coupling and
  Polarization in Molecular Transport Junctions: Beyond Tunneling to Function.
  \emph{Science} \textbf{2008}, \emph{319}, 1056--1060\relax
\mciteBstWouldAddEndPuncttrue
\mciteSetBstMidEndSepPunct{\mcitedefaultmidpunct}
{\mcitedefaultendpunct}{\mcitedefaultseppunct}\relax
\EndOfBibitem
\bibitem[Jean \latin{et~al.}(1992)Jean, Friesner, and
  Fleming]{jean1992application}
Jean,~J.~M.; Friesner,~R.~A.; Fleming,~G.~R. Application of a multilevel
  Redfield theory to electron transfer in condensed phases. \emph{The Journal
  of chemical physics} \textbf{1992}, \emph{96}, 5827--5842\relax
\mciteBstWouldAddEndPuncttrue
\mciteSetBstMidEndSepPunct{\mcitedefaultmidpunct}
{\mcitedefaultendpunct}{\mcitedefaultseppunct}\relax
\EndOfBibitem
\bibitem[Leegwater \latin{et~al.}(1997)Leegwater, Durrant, and
  Klug]{Leegwater:1997aa}
Leegwater,~J.~A.; Durrant,~J.~R.; Klug,~D.~R. Exciton Equilibration Induced by
  Phonons: Theory and Application to PS II Reaction Centers. \emph{The Journal
  of Physical Chemistry B} \textbf{1997}, \emph{101}, 7205--7210\relax
\mciteBstWouldAddEndPuncttrue
\mciteSetBstMidEndSepPunct{\mcitedefaultmidpunct}
{\mcitedefaultendpunct}{\mcitedefaultseppunct}\relax
\EndOfBibitem
\bibitem[Jeske \latin{et~al.}(2015)Jeske, Ing, Plenio, Huelga, and
  Cole]{Jeske:2015aa}
Jeske,~J.; Ing,~D.~J.; Plenio,~M.~B.; Huelga,~S.~F.; Cole,~J.~H. Bloch-Redfield
  equations for modeling light-harvesting complexes. \emph{The Journal of
  Chemical Physics} \textbf{2015}, \emph{142}, 064104\relax
\mciteBstWouldAddEndPuncttrue
\mciteSetBstMidEndSepPunct{\mcitedefaultmidpunct}
{\mcitedefaultendpunct}{\mcitedefaultseppunct}\relax
\EndOfBibitem
\bibitem[Storm \latin{et~al.}(2019)Storm, Rasmussen, Mikkelsen, and
  Hansen]{Storm:2019aa}
Storm,~F.~E.; Rasmussen,~M.~H.; Mikkelsen,~K.~V.; Hansen,~T. Computational
  construction of the electronic Hamiltonian for photoinduced electron transfer
  and Redfield propagation. \emph{Physical Chemistry Chemical Physics}
  \textbf{2019}, \emph{21}, 17366--17377\relax
\mciteBstWouldAddEndPuncttrue
\mciteSetBstMidEndSepPunct{\mcitedefaultmidpunct}
{\mcitedefaultendpunct}{\mcitedefaultseppunct}\relax
\EndOfBibitem
\bibitem[Pedersen \latin{et~al.}(2022)Pedersen, Rasmussen, and
  Mikkelsen]{Pedersen_2022_redfield_propagation_photoinduced_ET}
Pedersen,~J.; Rasmussen,~M.~H.; Mikkelsen,~K.~V. Redfield Propagation of
  Photoinduced Electron Transfer Reactions in Vacuum and Solution.
  \emph{Journal of Chemical Theory and Computation} \textbf{2022}, \emph{18},
  7052--7072\relax
\mciteBstWouldAddEndPuncttrue
\mciteSetBstMidEndSepPunct{\mcitedefaultmidpunct}
{\mcitedefaultendpunct}{\mcitedefaultseppunct}\relax
\EndOfBibitem
\bibitem[Weiss(2012)]{weiss_2012_quantum_dissipative_systems}
Weiss,~U. \emph{Quantum Dissipative Systems}; Series in modern condensed matter
  physics; World Scientific, 2012\relax
\mciteBstWouldAddEndPuncttrue
\mciteSetBstMidEndSepPunct{\mcitedefaultmidpunct}
{\mcitedefaultendpunct}{\mcitedefaultseppunct}\relax
\EndOfBibitem
\bibitem[Blum(2012)]{blum2012density}
Blum,~K. \emph{Density matrix theory and applications}; Springer Science \&
  Business Media, 2012; Vol.~64\relax
\mciteBstWouldAddEndPuncttrue
\mciteSetBstMidEndSepPunct{\mcitedefaultmidpunct}
{\mcitedefaultendpunct}{\mcitedefaultseppunct}\relax
\EndOfBibitem
\bibitem[The(2013)]{TheConceptofDecoherence_2013}
\emph{Molecular Excitation Dynamics and Relaxation}; John Wiley \& Sons, Ltd,
  2013; p 133–160\relax
\mciteBstWouldAddEndPuncttrue
\mciteSetBstMidEndSepPunct{\mcitedefaultmidpunct}
{\mcitedefaultendpunct}{\mcitedefaultseppunct}\relax
\EndOfBibitem
\bibitem[Nakajima(1958)]{nakajima1958quantum}
Nakajima,~S. On quantum theory of transport phenomena: Steady diffusion.
  \emph{Progress of Theoretical Physics} \textbf{1958}, \emph{20},
  948--959\relax
\mciteBstWouldAddEndPuncttrue
\mciteSetBstMidEndSepPunct{\mcitedefaultmidpunct}
{\mcitedefaultendpunct}{\mcitedefaultseppunct}\relax
\EndOfBibitem
\bibitem[Zwanzig(2004)]{Zwanzig_1960_projection_operators}
Zwanzig,~R. {Ensemble Method in the Theory of Irreversibility}. \emph{The
  Journal of Chemical Physics} \textbf{2004}, \emph{33}, 1338--1341\relax
\mciteBstWouldAddEndPuncttrue
\mciteSetBstMidEndSepPunct{\mcitedefaultmidpunct}
{\mcitedefaultendpunct}{\mcitedefaultseppunct}\relax
\EndOfBibitem
\bibitem[Lindblad(1976)]{lindblad1976generators}
Lindblad,~G. On the generators of quantum dynamical semigroups.
  \emph{Communications in Mathematical Physics} \textbf{1976}, \emph{48},
  119--130\relax
\mciteBstWouldAddEndPuncttrue
\mciteSetBstMidEndSepPunct{\mcitedefaultmidpunct}
{\mcitedefaultendpunct}{\mcitedefaultseppunct}\relax
\EndOfBibitem
\bibitem[Gorini \latin{et~al.}(1976)Gorini, Kossakowski, and
  Sudarshan]{gorini1976completely}
Gorini,~V.; Kossakowski,~A.; Sudarshan,~E. C.~G. Completely positive dynamical
  semigroups of N-level systems. \emph{Journal of Mathematical Physics}
  \textbf{1976}, \emph{17}, 821--825\relax
\mciteBstWouldAddEndPuncttrue
\mciteSetBstMidEndSepPunct{\mcitedefaultmidpunct}
{\mcitedefaultendpunct}{\mcitedefaultseppunct}\relax
\EndOfBibitem
\bibitem[Manzano(2020)]{manzano2020short}
Manzano,~D. A short introduction to the Lindblad master equation. \emph{AIP
  Advances} \textbf{2020}, \emph{10}, 025106\relax
\mciteBstWouldAddEndPuncttrue
\mciteSetBstMidEndSepPunct{\mcitedefaultmidpunct}
{\mcitedefaultendpunct}{\mcitedefaultseppunct}\relax
\EndOfBibitem
\bibitem[Hartmann and Strunz(2020)Hartmann, and Strunz]{hartmann2020accuracy}
Hartmann,~R.; Strunz,~W.~T. Accuracy assessment of perturbative master
  equations: Embracing nonpositivity. \emph{Physical Review A} \textbf{2020},
  \emph{101}, 012103\relax
\mciteBstWouldAddEndPuncttrue
\mciteSetBstMidEndSepPunct{\mcitedefaultmidpunct}
{\mcitedefaultendpunct}{\mcitedefaultseppunct}\relax
\EndOfBibitem
\bibitem[Tupkary \latin{et~al.}(2022)Tupkary, Dhar, Kulkarni, and
  Purkayastha]{tupkari_2022_limitations_of_lindblad_equations}
Tupkary,~D.; Dhar,~A.; Kulkarni,~M.; Purkayastha,~A. Fundamental limitations in
  Lindblad descriptions of systems weakly coupled to baths. \emph{Phys. Rev. A}
  \textbf{2022}, \emph{105}, 032208\relax
\mciteBstWouldAddEndPuncttrue
\mciteSetBstMidEndSepPunct{\mcitedefaultmidpunct}
{\mcitedefaultendpunct}{\mcitedefaultseppunct}\relax
\EndOfBibitem
\bibitem[Mukamel(1995)]{mukamel1995principles}
Mukamel,~S. \emph{Principles of Nonlinear Optical Spectroscopy}; Oxford series
  in optical and imaging sciences; Oxford University Press, 1995\relax
\mciteBstWouldAddEndPuncttrue
\mciteSetBstMidEndSepPunct{\mcitedefaultmidpunct}
{\mcitedefaultendpunct}{\mcitedefaultseppunct}\relax
\EndOfBibitem
\bibitem[Campaioli \latin{et~al.}(2024)Campaioli, Cole, and
  Hapuarachchi]{campaioli2024quantum}
Campaioli,~F.; Cole,~J.~H.; Hapuarachchi,~H. Quantum Master Equations: Tips and
  Tricks for Quantum Optics, Quantum Computing, and Beyond. \emph{PRX Quantum}
  \textbf{2024}, \emph{5}, 020202\relax
\mciteBstWouldAddEndPuncttrue
\mciteSetBstMidEndSepPunct{\mcitedefaultmidpunct}
{\mcitedefaultendpunct}{\mcitedefaultseppunct}\relax
\EndOfBibitem
\bibitem[Novoderezhkin \latin{et~al.}(2004)Novoderezhkin, Yakovlev, van
  Grondelle, and Shuvalov]{novoderezhkin2004coherent}
Novoderezhkin,~V.~I.; Yakovlev,~A.~G.; van Grondelle,~R.; Shuvalov,~V.~A.
  Coherent nuclear and electronic dynamics in primary charge separation in
  photosynthetic reaction centers: a Redfield theory approach. \emph{The
  Journal of Physical Chemistry B} \textbf{2004}, \emph{108}, 7445--7457\relax
\mciteBstWouldAddEndPuncttrue
\mciteSetBstMidEndSepPunct{\mcitedefaultmidpunct}
{\mcitedefaultendpunct}{\mcitedefaultseppunct}\relax
\EndOfBibitem
\bibitem[Kilin \latin{et~al.}(2000)Kilin, Kleinekath{\"o}fer, and
  Schreiber]{kilin2000electron}
Kilin,~D.; Kleinekath{\"o}fer,~U.; Schreiber,~M. Electron transfer in porphyrin
  complexes in different solvents. \emph{The Journal of Physical Chemistry A}
  \textbf{2000}, \emph{104}, 5413--5421\relax
\mciteBstWouldAddEndPuncttrue
\mciteSetBstMidEndSepPunct{\mcitedefaultmidpunct}
{\mcitedefaultendpunct}{\mcitedefaultseppunct}\relax
\EndOfBibitem
\bibitem[Novoderezhkin \latin{et~al.}(2007)Novoderezhkin, Dekker, and
  Van~Grondelle]{novoderezhkin2007mixing}
Novoderezhkin,~V.~I.; Dekker,~J.~P.; Van~Grondelle,~R. Mixing of exciton and
  charge-transfer states in photosystem II reaction centers: modeling of stark
  spectra with modified redfield theory. \emph{Biophysical journal}
  \textbf{2007}, \emph{93}, 1293--1311\relax
\mciteBstWouldAddEndPuncttrue
\mciteSetBstMidEndSepPunct{\mcitedefaultmidpunct}
{\mcitedefaultendpunct}{\mcitedefaultseppunct}\relax
\EndOfBibitem
\bibitem[Dzhioev and Kosov(2011)Dzhioev, and Kosov]{dzhioev2011super}
Dzhioev,~A.~A.; Kosov,~D.~S. Super-fermion representation of quantum kinetic
  equations for the electron transport problem. \emph{The Journal of Chemical
  Physics} \textbf{2011}, \emph{134}, 044121\relax
\mciteBstWouldAddEndPuncttrue
\mciteSetBstMidEndSepPunct{\mcitedefaultmidpunct}
{\mcitedefaultendpunct}{\mcitedefaultseppunct}\relax
\EndOfBibitem
\bibitem[Shavitt(1977)]{Shavitt1977}
Shavitt,~I. In \emph{Methods of Electronic Structure Theory}; Schaefer,~H.~F.,
  Ed.; Springer US: Boston, MA, 1977; pp 189--275\relax
\mciteBstWouldAddEndPuncttrue
\mciteSetBstMidEndSepPunct{\mcitedefaultmidpunct}
{\mcitedefaultendpunct}{\mcitedefaultseppunct}\relax
\EndOfBibitem
\bibitem[Helgaker \latin{et~al.}(2014)Helgaker, Jorgensen, and
  Olsen]{helgaker2014_mest}
Helgaker,~T.; Jorgensen,~P.; Olsen,~J. \emph{Molecular electronic-structure
  theory}; John Wiley \& Sons, 2014\relax
\mciteBstWouldAddEndPuncttrue
\mciteSetBstMidEndSepPunct{\mcitedefaultmidpunct}
{\mcitedefaultendpunct}{\mcitedefaultseppunct}\relax
\EndOfBibitem
\bibitem[Bogolyubov(1959)]{bogolyubov1959compensation}
Bogolyubov,~N.~N. The Compensation Principle and the Self-Consistent Field
  Method. \emph{Soviet Physics Uspekhi} \textbf{1959}, \emph{2}, 236--254\relax
\mciteBstWouldAddEndPuncttrue
\mciteSetBstMidEndSepPunct{\mcitedefaultmidpunct}
{\mcitedefaultendpunct}{\mcitedefaultseppunct}\relax
\EndOfBibitem
\bibitem[Slater \latin{et~al.}(1969)Slater, Mann, Wilson, and
  Wood]{slater_1969_hhf}
Slater,~J.; Mann,~J.; Wilson,~T.; Wood,~J. Nonintegral occupation numbers in
  transition atoms in crystals. \emph{Physical Review} \textbf{1969},
  \emph{184}, 672--694\relax
\mciteBstWouldAddEndPuncttrue
\mciteSetBstMidEndSepPunct{\mcitedefaultmidpunct}
{\mcitedefaultendpunct}{\mcitedefaultseppunct}\relax
\EndOfBibitem
\bibitem[Abdulnur \latin{et~al.}(1972)Abdulnur, Linderberg, {\"O}hrn, and
  Thulstrup]{abdulnur_1972_gchf}
Abdulnur,~S.~F.; Linderberg,~J.; {\"O}hrn,~U.; Thulstrup,~P.~W. Atomic
  central-field models for open shells with application to transition metals.
  \emph{Physical Review A} \textbf{1972}, \emph{6}, 889--898\relax
\mciteBstWouldAddEndPuncttrue
\mciteSetBstMidEndSepPunct{\mcitedefaultmidpunct}
{\mcitedefaultendpunct}{\mcitedefaultseppunct}\relax
\EndOfBibitem
\bibitem[Zerner(1989)]{zerner_1989_cahf}
Zerner,~M.~C. A configuration-averaged Hartree--Fock procedure.
  \emph{International journal of quantum chemistry} \textbf{1989}, \emph{35},
  567--575\relax
\mciteBstWouldAddEndPuncttrue
\mciteSetBstMidEndSepPunct{\mcitedefaultmidpunct}
{\mcitedefaultendpunct}{\mcitedefaultseppunct}\relax
\EndOfBibitem
\bibitem[Perdew \latin{et~al.}(1982)Perdew, Parr, Levy, and
  Balduz~Jr]{perdew_1982_fons_dft}
Perdew,~J.~P.; Parr,~R.~G.; Levy,~M.; Balduz~Jr,~J.~L. Density-functional
  theory for fractional particle number: Derivative discontinuities of the
  energy. \emph{Physical Review Letters} \textbf{1982}, \emph{49}, 1691\relax
\mciteBstWouldAddEndPuncttrue
\mciteSetBstMidEndSepPunct{\mcitedefaultmidpunct}
{\mcitedefaultendpunct}{\mcitedefaultseppunct}\relax
\EndOfBibitem
\bibitem[Ullrich and Kohn(2001)Ullrich, and Kohn]{ullrich_2001_ensembleDFT}
Ullrich,~C.; Kohn,~W. Kohn-Sham theory for ground-state ensembles.
  \emph{Physical Review Letters} \textbf{2001}, \emph{87}, 093001\relax
\mciteBstWouldAddEndPuncttrue
\mciteSetBstMidEndSepPunct{\mcitedefaultmidpunct}
{\mcitedefaultendpunct}{\mcitedefaultseppunct}\relax
\EndOfBibitem
\bibitem[Lee \latin{et~al.}(2022)Lee, Park, Nakata, Filatov, and
  Choi]{lee_2022_recent_advanced_in_edft}
Lee,~S.; Park,~W.; Nakata,~H.; Filatov,~M.; Choi,~C.~H. Recent advances in
  ensemble density functional theory and linear response theory for strong
  correlation. \emph{Bulletin of the Korean Chemical Society} \textbf{2022},
  \emph{43}, 17--34\relax
\mciteBstWouldAddEndPuncttrue
\mciteSetBstMidEndSepPunct{\mcitedefaultmidpunct}
{\mcitedefaultendpunct}{\mcitedefaultseppunct}\relax
\EndOfBibitem
\bibitem[Mussard and Toulouse(2017)Mussard, and Toulouse]{mussard_2017_fon_dft}
Mussard,~B.; Toulouse,~J. Fractional-charge and fractional-spin errors in
  range-separated density-functional theory. \emph{Molecular Physics}
  \textbf{2017}, \emph{115}, 161--173\relax
\mciteBstWouldAddEndPuncttrue
\mciteSetBstMidEndSepPunct{\mcitedefaultmidpunct}
{\mcitedefaultendpunct}{\mcitedefaultseppunct}\relax
\EndOfBibitem
\bibitem[Matveeva \latin{et~al.}(2023)Matveeva, Folkestad, and
  Høyvik]{Matveeva_2023_PBHF}
Matveeva,~R.; Folkestad,~S.~D.; Høyvik,~I.-M. Particle-Breaking Hartree–Fock
  Theory for Open Molecular Systems. \emph{The Journal of Physical Chemistry A}
  \textbf{2023}, \emph{127}, 1329--1341\relax
\mciteBstWouldAddEndPuncttrue
\mciteSetBstMidEndSepPunct{\mcitedefaultmidpunct}
{\mcitedefaultendpunct}{\mcitedefaultseppunct}\relax
\EndOfBibitem
\bibitem[Matveeva \latin{et~al.}(2024)Matveeva, Folkestad, Sannes, and
  H{\o}yvik]{matveeva2024particle}
Matveeva,~R. P.~N.; Folkestad,~S.~D.; Sannes,~B.~S.; H{\o}yvik,~I.-M.
  Particle-Breaking Unrestricted Hartree-Fock Theory for Open Molecular
  Systems. \emph{The journal of physical chemistry. A} \textbf{2024},
  \emph{128}, 1533--1542\relax
\mciteBstWouldAddEndPuncttrue
\mciteSetBstMidEndSepPunct{\mcitedefaultmidpunct}
{\mcitedefaultendpunct}{\mcitedefaultseppunct}\relax
\EndOfBibitem
\bibitem[H{\o}yvik and J{\o}rgensen(2016)H{\o}yvik, and
  J{\o}rgensen]{hoyvik2016characterization}
H{\o}yvik,~I.-M.; J{\o}rgensen,~P. Characterization and generation of local
  occupied and virtual Hartree--Fock orbitals. \emph{Chemical Reviews}
  \textbf{2016}, \emph{116}, 3306--3327\relax
\mciteBstWouldAddEndPuncttrue
\mciteSetBstMidEndSepPunct{\mcitedefaultmidpunct}
{\mcitedefaultendpunct}{\mcitedefaultseppunct}\relax
\EndOfBibitem
\bibitem[Jeske \latin{et~al.}(2012)Jeske, Cole, M{\"u}ller, Marthaler, and
  Sch{\"o}n]{jeske2012dual}
Jeske,~J.; Cole,~J.~H.; M{\"u}ller,~C.; Marthaler,~M.; Sch{\"o}n,~G. Dual-probe
  decoherence microscopy: probing pockets of coherence in a decohering
  environment. \emph{New Journal of Physics} \textbf{2012}, \emph{14},
  023013\relax
\mciteBstWouldAddEndPuncttrue
\mciteSetBstMidEndSepPunct{\mcitedefaultmidpunct}
{\mcitedefaultendpunct}{\mcitedefaultseppunct}\relax
\EndOfBibitem
\bibitem[Pauli(1928)]{pauli1928mastereq}
Pauli,~W. {\"U}ber das H-Theorem vom Anwachsen der Entropie vom Standpunkt der
  neuen Quantenmechanik. \emph{Probleme der Modernen Physik. Arnold Sommerfeld
  zum 60. Geburtstag gewindet von seinen Sch{\"u}lern} \textbf{1928},
  30--45\relax
\mciteBstWouldAddEndPuncttrue
\mciteSetBstMidEndSepPunct{\mcitedefaultmidpunct}
{\mcitedefaultendpunct}{\mcitedefaultseppunct}\relax
\EndOfBibitem
\bibitem[Berglund and Spicer(1964)Berglund, and
  Spicer]{berglund1964photoemission}
Berglund,~C.; Spicer,~W. Photoemission studies of copper and silver:
  Experiment. \emph{Physical Review} \textbf{1964}, \emph{136}, A1044\relax
\mciteBstWouldAddEndPuncttrue
\mciteSetBstMidEndSepPunct{\mcitedefaultmidpunct}
{\mcitedefaultendpunct}{\mcitedefaultseppunct}\relax
\EndOfBibitem
\bibitem[Reinert and H{\"u}fner(2005)Reinert, and
  H{\"u}fner]{reinert2005photoemission}
Reinert,~F.; H{\"u}fner,~S. Photoemission spectroscopy—from early days to
  recent applications. \emph{New Journal of Physics} \textbf{2005}, \emph{7},
  97\relax
\mciteBstWouldAddEndPuncttrue
\mciteSetBstMidEndSepPunct{\mcitedefaultmidpunct}
{\mcitedefaultendpunct}{\mcitedefaultseppunct}\relax
\EndOfBibitem
\bibitem[Verzijl \latin{et~al.}(2013)Verzijl, Seldenthuis, and
  Thijssen]{verzijl2013applicability}
Verzijl,~C. J.~O.; Seldenthuis,~J.~S.; Thijssen,~J.~M. Applicability of the
  wide-band limit in DFT-based molecular transport calculations. \emph{The
  Journal of Chemical Physics} \textbf{2013}, \emph{138}, 094102\relax
\mciteBstWouldAddEndPuncttrue
\mciteSetBstMidEndSepPunct{\mcitedefaultmidpunct}
{\mcitedefaultendpunct}{\mcitedefaultseppunct}\relax
\EndOfBibitem
\bibitem[Smith \latin{et~al.}(1974)Smith, Wertheim, H{\"u}fner, and
  Traum]{smith1974photoemission}
Smith,~N.~V.; Wertheim,~G.; H{\"u}fner,~S.; Traum,~M.~M. Photoemission spectra
  and band structures of d-band metals. IV. X-ray photoemission spectra and
  densities of states in Rh, Pd, Ag, Ir, Pt, and Au. \emph{Physical Review B}
  \textbf{1974}, \emph{10}, 3197\relax
\mciteBstWouldAddEndPuncttrue
\mciteSetBstMidEndSepPunct{\mcitedefaultmidpunct}
{\mcitedefaultendpunct}{\mcitedefaultseppunct}\relax
\EndOfBibitem
\bibitem[Rotter(2009)]{rotter2009non}
Rotter,~I. A non-Hermitian Hamilton operator and the physics of open quantum
  systems. \emph{Journal of Physics A: Mathematical and Theoretical}
  \textbf{2009}, \emph{42}, 153001\relax
\mciteBstWouldAddEndPuncttrue
\mciteSetBstMidEndSepPunct{\mcitedefaultmidpunct}
{\mcitedefaultendpunct}{\mcitedefaultseppunct}\relax
\EndOfBibitem
\bibitem[Nordlander and Tully(1990)Nordlander, and Tully]{nordlander1990energy}
Nordlander,~P.; Tully,~J. Energy shifts and broadening of atomic levels near
  metal surfaces. \emph{Physical Review B} \textbf{1990}, \emph{42}, 5564\relax
\mciteBstWouldAddEndPuncttrue
\mciteSetBstMidEndSepPunct{\mcitedefaultmidpunct}
{\mcitedefaultendpunct}{\mcitedefaultseppunct}\relax
\EndOfBibitem
\bibitem[Orr and Ward(1971)Orr, and Ward]{orr1971perturbation}
Orr,~B.; Ward,~J. Perturbation theory of the non-linear optical polarization of
  an isolated system. \emph{Molecular Physics} \textbf{1971}, \emph{20},
  513--526\relax
\mciteBstWouldAddEndPuncttrue
\mciteSetBstMidEndSepPunct{\mcitedefaultmidpunct}
{\mcitedefaultendpunct}{\mcitedefaultseppunct}\relax
\EndOfBibitem
\bibitem[Norman \latin{et~al.}(2005)Norman, Bishop, Jensen, and
  Oddershede]{norman2005nonlinear}
Norman,~P.; Bishop,~D.~M.; Jensen,~H. J.~A.; Oddershede,~J. Nonlinear response
  theory with relaxation: The first-order hyperpolarizability. \emph{The
  Journal of Chemical Physics} \textbf{2005}, \emph{123}, 194103\relax
\mciteBstWouldAddEndPuncttrue
\mciteSetBstMidEndSepPunct{\mcitedefaultmidpunct}
{\mcitedefaultendpunct}{\mcitedefaultseppunct}\relax
\EndOfBibitem
\bibitem[Dunning~Jr(1989)]{dunning1989gaussian}
Dunning~Jr,~T.~H. Gaussian basis sets for use in correlated molecular
  calculations. I. The atoms boron through neon and hydrogen. \emph{The Journal
  of chemical physics} \textbf{1989}, \emph{90}, 1007--1023\relax
\mciteBstWouldAddEndPuncttrue
\mciteSetBstMidEndSepPunct{\mcitedefaultmidpunct}
{\mcitedefaultendpunct}{\mcitedefaultseppunct}\relax
\EndOfBibitem
\bibitem[Liang \latin{et~al.}(2019)Liang, Liu, Chen, Qi, Kumar, Peera, Liu, He,
  and Liang]{liang2019oxygen}
Liang,~Z.; Liu,~C.; Chen,~M.; Qi,~X.; Kumar,~P.; Peera,~S.~G.; Liu,~J.; He,~J.;
  Liang,~T. Oxygen reduction reaction mechanism on P, N co-doped graphene: a
  density functional theory study. \emph{New Journal of Chemistry}
  \textbf{2019}, \emph{43}, 19308--19317\relax
\mciteBstWouldAddEndPuncttrue
\mciteSetBstMidEndSepPunct{\mcitedefaultmidpunct}
{\mcitedefaultendpunct}{\mcitedefaultseppunct}\relax
\EndOfBibitem
\bibitem[Griffiths and Higham(2010)Griffiths, and Higham]{Griffiths2010}
Griffiths,~D.~F.; Higham,~D.~J. \emph{Numerical Methods for Ordinary
  Differential Equations: Initial Value Problems}; Springer London: London,
  2010; pp 19--31\relax
\mciteBstWouldAddEndPuncttrue
\mciteSetBstMidEndSepPunct{\mcitedefaultmidpunct}
{\mcitedefaultendpunct}{\mcitedefaultseppunct}\relax
\EndOfBibitem
\bibitem[Breuer \latin{et~al.}(2016)Breuer, Laine, Piilo, and
  Vacchini]{breuer2016colloquium}
Breuer,~H.-P.; Laine,~E.-M.; Piilo,~J.; Vacchini,~B. Colloquium: Non-Markovian
  dynamics in open quantum systems. \emph{Reviews of Modern Physics}
  \textbf{2016}, \emph{88}, 021002\relax
\mciteBstWouldAddEndPuncttrue
\mciteSetBstMidEndSepPunct{\mcitedefaultmidpunct}
{\mcitedefaultendpunct}{\mcitedefaultseppunct}\relax
\EndOfBibitem
\bibitem[De~Vega and Alonso(2017)De~Vega, and Alonso]{de2017dynamics}
De~Vega,~I.; Alonso,~D. Dynamics of non-Markovian open quantum systems.
  \emph{Reviews of Modern Physics} \textbf{2017}, \emph{89}, 015001\relax
\mciteBstWouldAddEndPuncttrue
\mciteSetBstMidEndSepPunct{\mcitedefaultmidpunct}
{\mcitedefaultendpunct}{\mcitedefaultseppunct}\relax
\EndOfBibitem
\bibitem[Zhang(2019)]{zhang2019exact}
Zhang,~W.-M. Exact master equation and general non-Markovian dynamics in open
  quantum systems. \emph{The European Physical Journal Special Topics}
  \textbf{2019}, \emph{227}, 1849--1867\relax
\mciteBstWouldAddEndPuncttrue
\mciteSetBstMidEndSepPunct{\mcitedefaultmidpunct}
{\mcitedefaultendpunct}{\mcitedefaultseppunct}\relax
\EndOfBibitem
\bibitem[Tanimura and Kubo(1989)Tanimura, and Kubo]{tanimura_kubo1989HEOM}
Tanimura,~Y.; Kubo,~R. Time evolution of a quantum system in contact with a
  nearly Gaussian-Markoffian noise bath. \emph{Journal of the Physical Society
  of Japan} \textbf{1989}, \emph{58}, 101--114\relax
\mciteBstWouldAddEndPuncttrue
\mciteSetBstMidEndSepPunct{\mcitedefaultmidpunct}
{\mcitedefaultendpunct}{\mcitedefaultseppunct}\relax
\EndOfBibitem
\bibitem[Zheng \latin{et~al.}(2012)Zheng, Xu, Xu, Jin, Hu, and
  Yan]{zheng2012heom_rev}
Zheng,~X.; Xu,~R.; Xu,~J.; Jin,~J.; Hu,~J.; Yan,~Y. Hierarchical equations of
  motion for quantum dissipation and quantum transport. \emph{Progress in
  Chemistry} \textbf{2012}, \emph{24}, 1129\relax
\mciteBstWouldAddEndPuncttrue
\mciteSetBstMidEndSepPunct{\mcitedefaultmidpunct}
{\mcitedefaultendpunct}{\mcitedefaultseppunct}\relax
\EndOfBibitem
\bibitem[Wu and Van~Voorhis(2006)Wu, and Van~Voorhis]{wu2006extracting}
Wu,~Q.; Van~Voorhis,~T. Extracting electron transfer coupling elements from
  constrained density functional theory. \emph{The Journal of Chemical Physics}
  \textbf{2006}, \emph{125}, 164105\relax
\mciteBstWouldAddEndPuncttrue
\mciteSetBstMidEndSepPunct{\mcitedefaultmidpunct}
{\mcitedefaultendpunct}{\mcitedefaultseppunct}\relax
\EndOfBibitem
\bibitem[Kaduk \latin{et~al.}(2012)Kaduk, Kowalczyk, and
  Van~Voorhis]{kaduk2012constrained}
Kaduk,~B.; Kowalczyk,~T.; Van~Voorhis,~T. Constrained density functional
  theory. \emph{Chemical reviews} \textbf{2012}, \emph{112}, 321--370\relax
\mciteBstWouldAddEndPuncttrue
\mciteSetBstMidEndSepPunct{\mcitedefaultmidpunct}
{\mcitedefaultendpunct}{\mcitedefaultseppunct}\relax
\EndOfBibitem
\bibitem[Mulliken(1952)]{mulliken1952molecular}
Mulliken,~R.~S. Molecular compounds and their spectra. II. \emph{Journal of the
  American Chemical Society} \textbf{1952}, \emph{74}, 811--824\relax
\mciteBstWouldAddEndPuncttrue
\mciteSetBstMidEndSepPunct{\mcitedefaultmidpunct}
{\mcitedefaultendpunct}{\mcitedefaultseppunct}\relax
\EndOfBibitem
\bibitem[Hush(1961)]{hush1961adiabatic}
Hush,~N.~S. Adiabatic theory of outer sphere electron-transfer reactions in
  solution. \emph{Transactions of the Faraday Society} \textbf{1961},
  \emph{57}, 557--580\relax
\mciteBstWouldAddEndPuncttrue
\mciteSetBstMidEndSepPunct{\mcitedefaultmidpunct}
{\mcitedefaultendpunct}{\mcitedefaultseppunct}\relax
\EndOfBibitem
\bibitem[Hush(1968)]{hush1968homogeneous}
Hush,~N.~S. Homogeneous and heterogeneous optical and thermal electron
  transfer. \emph{Electrochimica Acta} \textbf{1968}, \emph{13},
  1005--1023\relax
\mciteBstWouldAddEndPuncttrue
\mciteSetBstMidEndSepPunct{\mcitedefaultmidpunct}
{\mcitedefaultendpunct}{\mcitedefaultseppunct}\relax
\EndOfBibitem
\bibitem[Cave and Newton(1996)Cave, and Newton]{cave1996generalization}
Cave,~R.~J.; Newton,~M.~D. Generalization of the Mulliken-Hush treatment for
  the calculation of electron transfer matrix elements. \emph{Chemical physics
  letters} \textbf{1996}, \emph{249}, 15--19\relax
\mciteBstWouldAddEndPuncttrue
\mciteSetBstMidEndSepPunct{\mcitedefaultmidpunct}
{\mcitedefaultendpunct}{\mcitedefaultseppunct}\relax
\EndOfBibitem
\end{mcitethebibliography}
